\documentclass[superscriptaddress,twocolumn]{revtex4}
\usepackage{mathrsfs}
\usepackage{amssymb}
\usepackage[tbtags]{amsmath}
\usepackage{graphicx}
\usepackage{epsfig,graphicx,times}
\usepackage{color}
\usepackage{subfigure}
\usepackage{setspace}
\usepackage{extarrows}
\usepackage[hidelinks,colorlinks=true,linkcolor=blue,citecolor=blue]{hyperref}

\begin{document}
\title{Quantum detection of millimeter wave electric fields with driving surface-state electrons}
\author{Miao Zhang~\footnote{zhangmiao@swjtu.edu.cn}}
\affiliation{School of Physical Science and Technology, Southwest Jiaotong University, Chengdu 610031, China}
\author{Y. F. Wang}
\affiliation{School of Physical Science and Technology, Southwest Jiaotong University, Chengdu 610031, China}
\author{X. Y. Peng}
\affiliation{School of Physical Science and Technology, Southwest Jiaotong University, Chengdu 610031, China}
\author{X. N. Feng}
\affiliation{Information Quantum Technology Laboratory, International Cooperation Research Center of China Communication and Sensor Networks for Modern Transportation, School of Information Science and Technology, Southwest Jiaotong University, Chengdu 610031, China}
\author{S. R. He}
\affiliation{Information Quantum Technology Laboratory, International Cooperation Research Center of China Communication and Sensor Networks for Modern Transportation, School of Information Science and Technology, Southwest Jiaotong University, Chengdu 610031, China}
\author{Y. F. Li}
\affiliation{Information Quantum Technology Laboratory, International Cooperation Research Center of China Communication and Sensor Networks for Modern Transportation, School of Information Science and Technology, Southwest Jiaotong University, Chengdu 610031, China}
\author{L. F. Wei~\footnote{lfwei@swjtu.edu.cn}}
\affiliation{Information Quantum Technology Laboratory, International Cooperation Research Center of China Communication and Sensor Networks for Modern Transportation, School of Information Science and Technology, Southwest Jiaotong University, Chengdu 610031, China}

\date{\today}

\begin{abstract}
We introduce a spin-based receiver to sensitively detect the electric fields of millimeter (mm) waves by using quantum interferometric approach. The proposed quantum sensor consists of many surface-state electrons trapped individually on liquid helium by an electrode-network at the bottom of the liquid helium film. A dc-current in this chip is biased to generate a strong spin-orbit coupling of each of the trapped electrons. The mm wave signals are conducted to non-dissipatedly drive the orbital motions of the trapped electrons and result in the Stark shifts of the spin-orbit dressed states of the electrons. As a consequence, the electric fields of the conducted mm waves could be detected sensitively by using the Hahn echo interferometry with the long-lived spin states of the electrons trapped on liquid helium.
\end{abstract}

\maketitle
\section{Introduction}
Quantum sensors make use of the quantum properties, such as states superposition and entanglement, to measure a physical quantity with high accuracy~\cite{Quantum sensing,Cold-atom-RMP}. An excellent quantum sensor exhibits two important qualities, (i) the long-lived superposition state which means the immunity to noises, and (ii) the strong couplings to some special signals which mean high sensitivity~\cite{PRB-while}.
In recent years, a series of physical systems, such as the trapped ions~\cite{trapped-ion-RF}, cold atoms clouds~\cite{Cold-atom-magentci,AC-Stark RMP}, solid-state spins~\cite{NV-RMP}, SQUID~\cite{RSI-SQUID}, and the Rydberg atoms~\cite{ShanXi-GHz}, have been implemented as the quantum sensors to detect the electromagnetic fields or gravity. Specially, the electromagnetic detections in the frequency-range from $0.1$ to $10$ terahertz (THz)~\cite{THz-NatureMaterials} have been paid much attention for various application fields, such as astronomical observation, biomedical diagnosis, and the THz radar~\cite{THz-NaturePhotonics}, etc.. Specifically, the Rydberg atoms are sensitive to a broad of electromagnetic waves~\cite{kHz-THz-JPB,Electrometry}, and thus can be developed for the THz sensing~\cite{THz-NPhoton,THz-NC,THz-PRX,THz-ZHU}.

In the present work, by using a conceptually new system, i.e., the spin of electron trapped on liquid helium~\cite{Lyon,Dykman2023,Schuster2010,MiaoPRB,Lyon-2011}, we investigate the quantum sensor of the electric fields of millimeter (mm) waves. Compared to the charge freedom of the electron trapped either on liquid helium~\cite{Science-qubit,Dykman-qubit,SchusterNC,Ising} or in Rydberg atoms, the electronic spin freedom has the longer coherent time up to millisecond (ms) within the low temperature at mK level~\cite{Schuster2010}.
Although the directly magnetic couplings between spin and the mm waves are significantly weak and thus almost negligible, we show that the indirect couplings between them can be strong sufficiently by using the spin-orbit coupling of the trapped electron on liquid helium, wherein the transition frequency between the orbital-levels of the surface-binding electrons is in the mm waves band.

The existence of surface-binding state of electron on liquid helium is due to two facts. First, there is a barrier at the inert helium liquid surface  about $1$ eV that prevents electrons from entering the liquid helium interior~\cite{Sommer1964}. Second, the electronic image charge inside liquid helium attracts the test charge moving toward the liquid helium surface. Therefore, electron on liquid helium serves as a one-dimensional (1D) hydrogen-like atom with the Bohr radius about $z_b\approx7.6$~nm~\cite{Grimes1974,RMP-Cole,Saturation}. The transition frequency between the lowest two levels of such a 1D atom is just in the mm wave band~\cite{zhangOL,Wang}. Compared to the usual Rydberg atoms, here the Bohr radius is not too small to lead a considerable electric dipole interaction between the 1D atom and the applied mm wave. Unusually, here exists
a linear Stark effect in the electric dipole interactions due to the symmetry breaking wave function of the 1D atom. Experimentally, this effect can be applied to realize the controllable couplings between atoms and the mm waves~\cite{DenisPRL2019,DenisJLTP2021,DenisPRL2021, DenisNJP2022,DenisJLTP2022,UK2023}. Probably, by manipulating the spin-orbit coupling of this symmetry breaking atom, the long-lived spin states of electron could be utilized as a quantum sensor for the mm wave electric fields.

To achieve this end, we design a scalable electrode-network to trap a series of electrons individually on superfluid helium film. In this chip, a dc-current is loaded to generate the local magnetic fields (with high gradients) for implementing the strong spin-orbit couplings of every trapped electron. Specifically, for the 1D atom introduced above, its spin-orbit coupling strength can reach up to the megahertz (MHz) level. With this coupling, the electric field of the applied mm wave can also affect the spins effectively, for example, result in the level-splitting of spins. We call this effect the spin ac-Stark shift and show that electronic orbital states are not required to be excited during the driving for this effect. Therefore, the present spin ac-Stark shift could be utilized to implement the sensitive detection of the applied mm waves, by using the Hahn echo interferometry (called also the spin echo interferometry)~\cite{Hahn-echo1,Hahn-echo2}, which has been successfully applied to precisely measure the static magnetic fields in the system of nitrogen-vacancy (NV) centers~\cite{NV-RMP}.

The article is organized as follows. In Sec. II, we design an electrode-network to trap a series of electrons individually on liquid helium. This electrode-network also allows us to load a dc-current for generating the strong spin-orbit coupling of every trapped electron. With the presence of spin-orbit coupling, in Sec. III, we show how the ac-Stark shift of electronic spin can be generated by using the electric dipole interaction between mm wave and the 1D atom introduced above. In Sec. IV, we show how the spins echo interferometry~\cite{NV-RMP} together with the technique of surface electrons detection~\cite{DenisPRL2019} can be applied to detect the present ac-Stark shift of electronic spin on liquid helium. With these techniques, the reachable sensitivity of the mm wave detection is then analyzed. Finally, we present our conclusion in Sec. V.

\begin{figure}
\subfigure[]{\includegraphics[width=7.5cm]{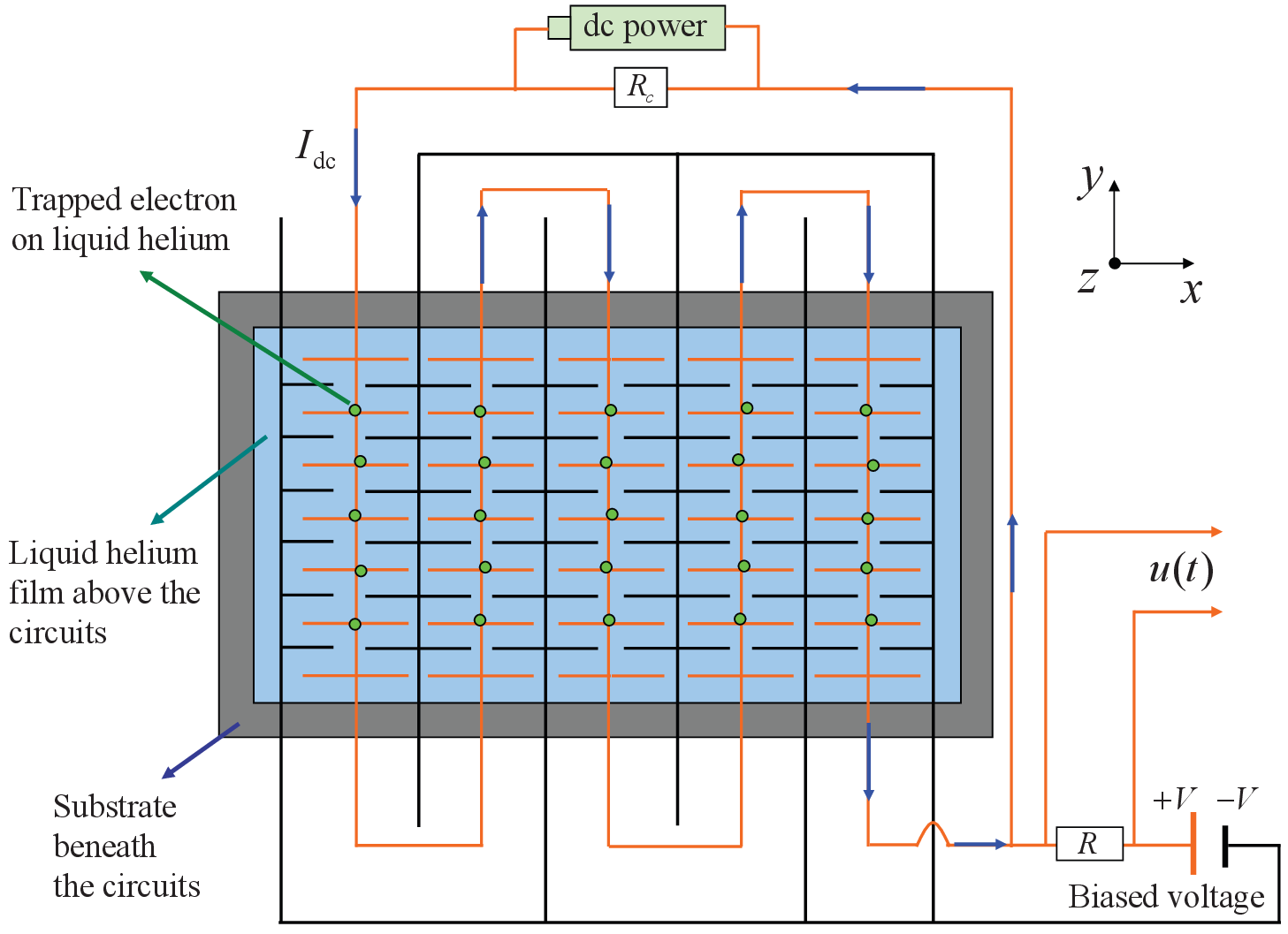}}
\subfigure[]{\includegraphics[width=6cm]{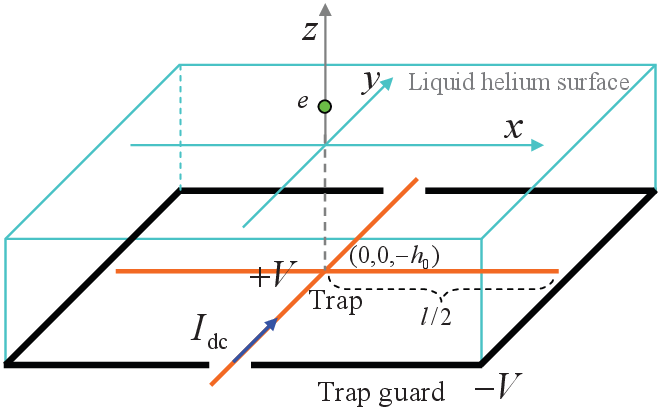}}
\caption{(a) An electrode-network for trapping a series of electrons individually on superfluid helium film, and where a dc-current $I_{\rm {dc}}$ on the orange wire is used to generate the spin qubits and also the strong spin-orbit couplings. With these couplings and the mm wave driving orbital motions, the information of spin qubits are transferred into the output voltage $u(t)$ for detection.
(b) A magnified cell of the designed spins chip, where the orange ``+'' electrode with the biased voltage
$+V$ is used to laterally trap the surface-state electron, and the black edging frame with voltage $-V$ serves as the trap guard.}
\label{fig1}
\end{figure}

\section{The micro-well arrays to trap a series of electrons individually on liquid Helium}
\begin{figure}[tbp]
  \subfigure[]{\includegraphics[width=5.5cm]{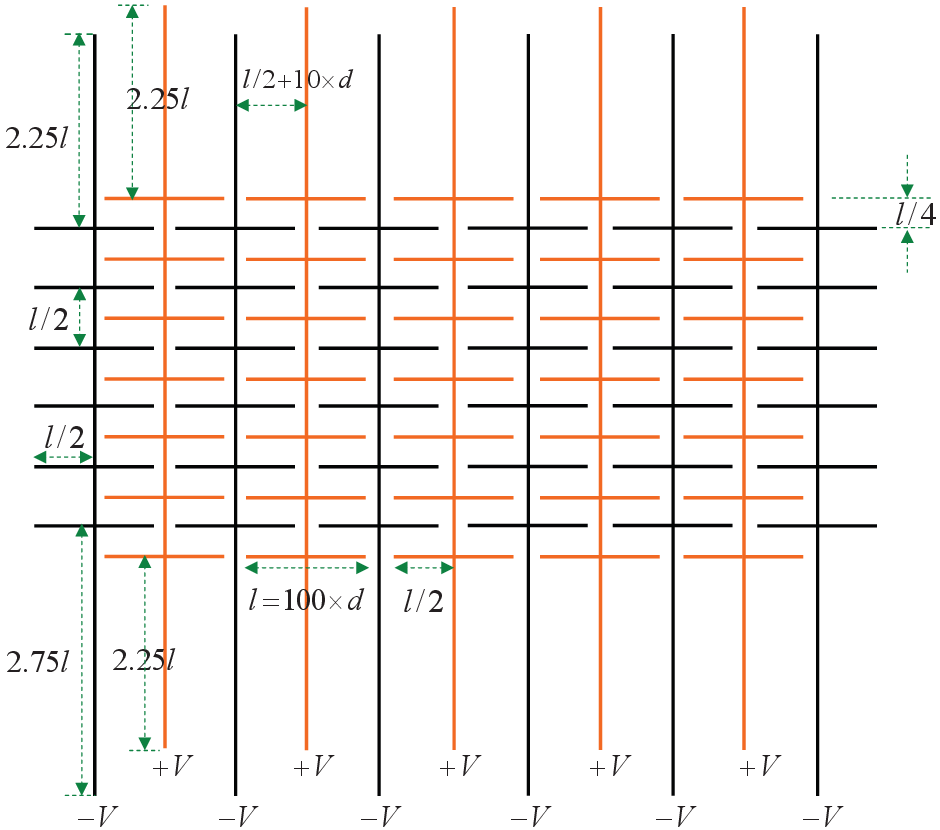}}
  \subfigure[]{\includegraphics[width=6.5cm]{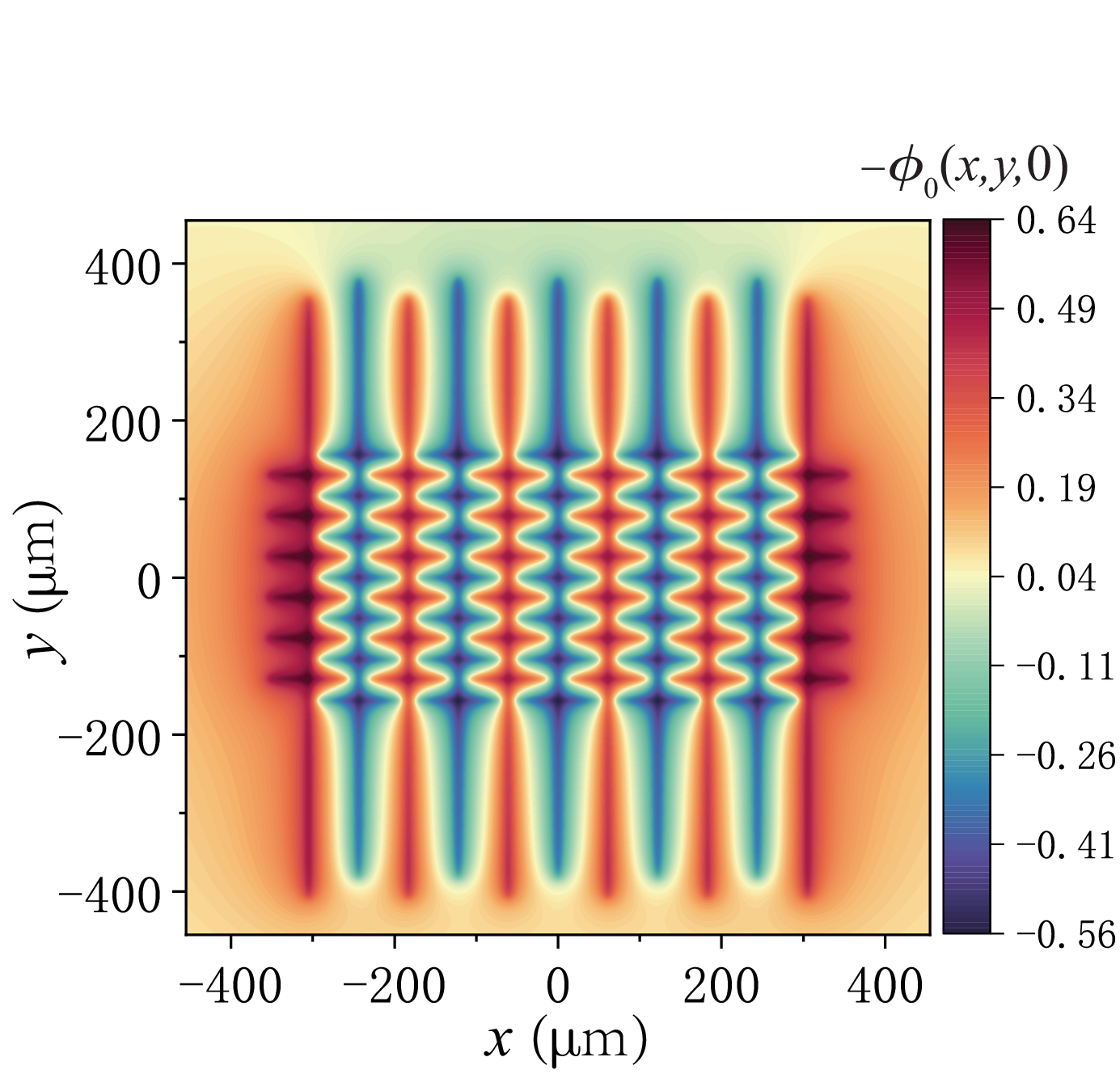}}
  \caption{(a) A model used to numerically compute the electrostatic potentials generated by the proposed electrode-network, wherein the solid lines are the physical electrodes, with the identical sectional area $d\times d=1\,\mu{\rm m}\times1\,\mu{\rm m}$ of the conductive wires. (b) The obtained solution $\phi_0(x,y,0)$ at $z=0$ plane, with the unit being an arbitrary voltage $V$. Throughout the paper, the Cartesian coordinate system is established on the surface of liquid helium, i.e., the metallic lattice is at the plane of $z=-h_0$, with $h_0=5\,\mu{\rm m}$ being the considered thickness of the liquid helium film.}
  \label{fig2}
\end{figure}

The prototype chip designed here is sketched in Fig.~\textcolor{blue}{1(a)}, which exhibits two important functions; trapping a series of electrons individually on liquid helium, and a dc-current can be loaded to generate the strong spin-orbit coupling of every trapped
electron. Specially, the first function, i.e., the micro-trap for one of the electrons, can be explained as follows. In the Cartesian coordinate system shown in Fig.~\textcolor{blue}{1(b)}, the electrostatic potential
\begin{equation}
\begin{aligned}
\phi_0(\textbf{r})
\approx&V_{0}+x E_x+y E_y+z E_z\\
&+
x^2 Q_{xx}+y^2 Q_{yy}+z^2 Q_{zz}\\
&+xy Q_{xy}+xz Q_{xz}+yz Q_{yz}\,,
\end{aligned}
\end{equation}
can be generated around the point $\textbf{r}=(x,y,z)\rightarrow(0,0,0)$ for trapping a single electron. Above, $V_{0}$, $E_i$, and $Q_{ij}$, with $i,j=x,y,z$, are the coefficients of the Taylor expansion.

Due to the symmetry, i.e., $\phi_0(x,y,z)=\phi_0(-x,y,z)$ and $\phi_0(x,y,z)=\phi_0(x,-y,z)$, we have $E_x=E_y=0$ and $Q_{xy}=Q_{xz}=Q_{yz}=0$. As a consequence, the above electrostatic potential reduces to
\begin{equation}
\begin{aligned}
\phi_0(\textbf{r})\approx V_{0}+z E_z +x^2 Q_{xx}+y^2Q_{yy}+z^2 Q_{zz}\,,
\end{aligned}
\end{equation}
which can trap the electron as a 2D harmonic oscillator in $x-y$ plane.
Using the geometric parameters in Fig.~\textcolor{blue}{2(a)}, a scalable traps-array is numerically showed by Fig.~\textcolor{blue}{2(b)}, wherein $E_{z}\approx 0.487\times 10^{3}V/{\rm cm}$,  $Q_{xx}\approx 0.394\times 10^{6}V/{\rm (cm)^2}$,  $Q_{yy}\approx 0.393\times 10^{6}V/{\rm (cm)^2}$, and $Q_{zz}\approx -0.798\times 10^{6}V/{\rm (cm)^2}$ for an arbitrary bias-voltage $V$.

Suppose that the number of potential wells is more than the number of electrons, and the depth of potentials is set at a level that is little larger than $k_bT/e$ in the experimental initial stage (with $k_b$, $e$, and $T$ being respectively Boltzmann constant, electronic charge, and the temperature). Therefore, if two or more electrons present within a single potential well, the effective potential well for one of the electrons is reduced due to the repulsion of the excess electrons in this zone. Consequently, particle's kinetic energy is larger its potential energy, and then the electrons spread into the other artificial potential wells. After the stage of permeation, we adiabatically enlarge the biased voltage (i.e., the depth of the potential wells), such that the trapped electron can be froze into its vibrational ground state via the spontaneous radiative emission. Different from the usual method for electrons crystallization on liquid helium~\cite{WignerPRL,WignerPL}, our artificial potential wells are separated distantly, and then the Coulomb interactions between the localized electrons are negligible.

Therefore, the Hamiltonian of one of the trapped electrons on helium ``substrate" can be written as
\begin{equation}
\begin{aligned}
H_0=&\frac{\textbf{p}^2}{2m_e}
-\frac{ee_{\rm eff}}{4\pi\epsilon_0z}+ezE_z\\
&+\frac{1}{2}m_e(\omega_{x}^2x^2+\omega_{y}^2y^2)+e z^2Q_{zz}\,.
\end{aligned}
\end{equation}
Here, $m_e$ is electronic mass, and $\epsilon_0$ the vacuum permittivity. Specially, $e_{\rm eff}=(\epsilon-1)e/[4(\epsilon+1)]$ is the image charge inside liquid helium, with $\epsilon\approx1.057$ being the dielectric constant of liquid helium~\cite{Dykman-qubit}. As we well known that, the $z$-directional vibration of the surface electron generates a 1D hydrogen-like atom~\cite{RMP-Cole}, even without the artificial trap. The lowest two levels, with transition frequency $\omega^{(0)}_a/(2\pi)\approx119$~GHz (corresponding to a wavelength of $2.5$~mm), can be regarded as an atomic qubit working within the mm wave band. Within the trap designed above, the transition frequency of the atomic qubit is changed to be $\omega_a/(2\pi)\approx160$~GHz, for $E_z\approx0.487\times10^3\,\,V/{\rm cm}$ and $V=0.1$ Volt.
While, the frequency of the lateral vibrations of the trapped electron in the $x-y$ plane is $\omega_{j}=\sqrt{2eQ_{jj}/m_e}\approx12$~GHz, with $j=x,y$.

The last term in Hamiltonian~({\color{blue}3}) is negligible as it refers to the square of Bohr radius $z_b\approx7.6$~nm~\cite{Dykman-qubit}. As a consequence, considering only the lowest two levels of the 1D hydrogen-like atom, we rewrite Hamiltonian ({\color{blue}3}) as
\begin{equation}
\begin{aligned}
\hat{H}_0=&\hbar\left(\omega_1|1\rangle\langle 1|
+\omega_2|2\rangle\langle 2|\right)\\
&
+\hbar\omega_{x}(\hat{a}_x^\dagger\hat{a}_x+\frac{1}{2})
+\hbar\omega_{y}(\hat{a}_y^\dagger\hat{a}_y+\frac{1}{2})\,,
\end{aligned}
\end{equation}
with $\omega_a=\omega_2-\omega_1\approx2\pi\times 160~{\rm GHz}$ being the transition frequency of atomic qubit mentioned above. Obviously, $\hbar\omega_1$ and $\hbar\omega_2$ are the energy-eigenvalues of the ground state $|1\rangle$ and the first excited one $|2\rangle$ of the 1D Hydrogen-like atom, respectively. In the second line of Hamiltonian ({\color{blue}4}), $\hat{a}_x^\dagger$ and $\hat{a}_x$ are the creation and annihilation operators of the $x$-directional harmonic oscillation, respectively. They are defined by $\hat{x}=x_{0}(\hat{a}_x^\dagger+\hat{a}_x)$ and $\hat{p}_x=ip_{x0}(\hat{a}_x^\dagger-\hat{a}_x)$ with the coefficients   $x_{0}=\sqrt{\hbar/(2m_e\omega_{x})}$ and $p_{x0}=\sqrt{\hbar m_e\omega_{x}/2}$.
The definitions of $y$-directional operators $\hat{a}_y^\dagger$ and $\hat{a}_y$
are similar.

\section{The spin-orbit couplings of a trapped electron}
In this section, we discuss how to implement the strong spin-orbit
coupling of an electron trapped on liquid helium, and how the ac-Stark effect of electronic spin can be generated via the electric dipole interaction between the mm wave and the 1D atom introduced above.

\subsection{The Hamiltonian of a trapped electron within the static magnetic field and the mm wave}
To achieve the spin-orbit coupling of the trapped electron, we now apply a dc-current $I_{\rm dc}$ to the chip.
In the Cartesian coordinate system shown in Fig.~\textcolor{blue}{1(b)}, the magnetic field generated by the $y$-directional dc-current beneath the liquid helium of depth $h_0$ can be expressed as
\begin{equation}
\begin{aligned}
\textbf{B}_0
&=\left(\frac{\mu_0I_{\rm dc}}{2\pi}\right)\frac{(h_0+z)\textbf{e}_x-x\textbf{e}_z}
{x^2+(h_0+z)^2}\\
&\approx B_0\left[(1-\frac{z}{h_0})\textbf{e}_x
-\frac{x}{h_0}\textbf{e}_z
+\mathcal{O}(1/h_0^2)\right]\,.
\end{aligned}
\end{equation}
Here, $B_0=\mu_0I_{\rm dc}/(2\pi h_0)$, with $\mu_0=4\pi\times10^{-7}\,\,{\rm T}\cdot{\rm m}/{\rm A}$ being the vacuum permeability. $\textbf{e}_x$ and $\textbf{e}_z$ denote the unit vectors along the $x$ and $z$ directions, respectively. We emphasize that the approximated solution in Eq.~({\color{blue}5}) does not violate the Maxwell's equations $\nabla\cdot\textbf{B}_0=0$ and $\nabla\times\textbf{B}_0=0$. Furthermore, nearing the coordinate origin, i.e., $(x,y,z)\approx(0,0,0)$, the magnetic field generated by the non-adjacent dc-currents on the chip are negligible, i.e., $\textbf{B}_0(x\pm l)
\approx\pm[\mu_0I_{\rm dc}/(2\pi l)]\textbf{e}_z$ with $h_0\ll l$, and therefore $\textbf{B}_0(x+l)+\textbf{B}_0(x-l)\approx0$.

According to Eq.~({\color{blue}5}), the magnetic vector potential reads
\begin{equation}
\begin{aligned}
\textbf{A}_0&=-\frac{h_0 B_0}{2}\ln\left[\frac{x^2}{h_0^2}
+\left(1+\frac{z}{h_0}\right)^2\right]\textbf{e}_y\\
&
\approx -B_0\left[z-\frac{z^2}{2 h_0}+\frac{x^2}{2 h_0}
+\mathcal{O}(1/h_0^2)\right]
\textbf{e}_y\,,
\end{aligned}
\end{equation}
with $\textbf{e}_y$ being the unit vector along $y$-direction. Obviously,
the above vector potential satisfies its definition
$\textbf{B}_0=\nabla\times \textbf{A}_0$ and also the Coulomb gauge $\nabla\cdot \textbf{A}_0=0$.

In addition to the dc-current, we assume that the mm wave is applied simultaneously, and therefore, the Hamiltonian for the non-relativistic motion of a trapped electron should be generally written as
\begin{equation}
\begin{aligned}
\hat{H}_{\rm PL}&=\frac{(\hat{\textbf{p}}-e\textbf{A})^2}{2m_e}+e\phi
+u_b\hat{\textbf{S}}\cdot\textbf{B}\,.
\end{aligned}
\end{equation}
Here, $\hat{\textbf{S}}$ is the Pauli operator of electronic spin, $u_b\approx9.3\times10^{-24}\,{\rm A}{\rm m}^2$ is Bohr magneton, and $\textbf{B}=\textbf{B}_0+\textbf{B}_w$ with $\textbf{B}_w$ being the oscillating magnetic field of the applied mm wave. Correspondingly, the magnetic vector potential in the above Hamiltonian
is expanded as $\textbf{A}=\textbf{A}_0+\textbf{A}_w$, with $\textbf{A}_0$ being the vector potential ({\color{blue}6}) generated by the static magnetic field, and $\textbf{A}_w$ the vector potential from the mm wave. Similarly, the electric scalar potential can be written as a form of $\phi=\Phi_0+\phi_w$.  Here, $\phi_w$ is due to the inputting mm wave, and $\Phi_0$ the previous electrostatic potential presented for trapping the electron on liquid helium.

For the mm wave electric field $\textbf{E}_{w}=\textbf{e}_zE_w\cos(\omega_T t-k_Tx)=-\partial_t\textbf{A}_w$ (with amplitude $E_w$, frequency $\omega_T$, and the wave number $k_T$), one can find that $\phi_w=0$ and $\textbf{A}_w=-(E_w\textbf{e}_z/\omega_T)\sin(\omega_T t-k_Tx)$ under the Coulomb gauge $\nabla\cdot\textbf{A}_w=0$. Performing the electric dipole approximation, see Appendix~\textcolor{blue}{A}, Hamiltonian ({\color{blue}7}) can be written as
\begin{equation}
\begin{aligned}
\hat{H}&\approx\hat{H}_0-e\textbf{r}\cdot\textbf{E}_w
+u_b\hat{\textbf{S}}\cdot\textbf{B}_0\,,
\end{aligned}
\end{equation}
with $\hat{H}_0$ that has already been given by Eq.~({\color{blue}4}). In atomic qubit space, i.e., $z=\sum_{n,m=1}^2z_{nm}|n\rangle\langle m|$ with $z_{nm}=\langle n|z|m\rangle$, the electric dipole interaction in Eq.~({\color{blue}8}) can be further written as
\begin{equation}
\begin{aligned}
e\textbf{r}\cdot\textbf{E}_w\approx\hbar\Omega_{12}\left(|1\rangle\langle 2|+{\rm h.c.}\right)
\cos(\omega_T t-k_Tx)\,,
\end{aligned}
\end{equation}
where $\Omega_{12}=eE_{w}z_{12}/\hbar$ is the atomic Rabi frequency of the
electric dipole interaction. The diagonal terms $ez_{nn}E_w\cos(\omega_Tt-k_Tx)|n\rangle\langle n|$ in the electric dipole interaction have been neglected, as $ez_{nn}E_w\ll\hbar \omega_T$ for the high-frequency driving.

With the magnetic field shown in~({\color{blue}5}), the spin-related term in Eq.~({\color{blue}8}) can be expanded as
\begin{equation}
\begin{aligned}
u_b\hat{\textbf{S}}\cdot\textbf{B}_0
\approx\frac{\hbar\omega_s}{2}\hat{S}_z
-\frac{\hbar\omega_s}{2}\frac{\hat{z}}{h_0}\hat{S}_z
-\frac{\hbar\omega_s}{2}\frac{\hat{x}}{h_0}\hat{S}_x\,.
\end{aligned}
\end{equation}
The first term, with $\hat{S}_z=|\uparrow\rangle\langle \uparrow|
-|\downarrow\rangle\langle\downarrow|$, defines a spin qubit with the transition frequency $\omega_s=2\mu_bB_0/\hbar$ between the Zeeman-splitting eigenstates $|\uparrow\rangle$ and $|\downarrow\rangle$. The second term describes a spin-dependent force along $z$ direction, which results in a spin-dependent Stark effect of the atomic qubit.
The third term, with $\hat{S}_x=|\uparrow\rangle\langle \downarrow|
+|\downarrow\rangle\langle\uparrow|$, refers to the coupling between spin and the lateral oscillation of the trapped electron.

Numerically, supposing $I_{\rm dc}=0.5$~A and $h_0=5\,\,\mu{\rm m}$~\cite{Schuster2010}, we have $\omega_{s}\approx3.5\,{\rm GHz}$ with $B_0=\mu_0I_{\rm dc}/(2\pi h)\approx0.01\,{\rm T}$, where $\mu_0=4\pi\times10^{-7}\,\,{\rm T}\cdot{\rm m}/{\rm A}$ is the vacuum permeability. For the materials with the critical current
density $10^8\,{\rm A}/{\rm cm}^2$~\cite{RSI-SQUID}, the considered current $I_{\rm dc}=0.5$~A on an electrode of sectional area $1\,\mu{\rm m}\times1\,\mu{\rm m}$ is feasible. In the low temperature such as $T=10$ mK~\cite{refrigerator,SchusterNature,Schuster2023,QCT}, the spin qubit with transition frequency $\omega_s\approx3.5\,{\rm GHz}$ can be initialized into its ground state as $\hbar\omega_s/k_b\approx26$~mK, with Boltzmann constant $k_b=1.38\times10^{-23}{\rm J}/{\rm K}$.

\subsection{The Rabi oscillation of atomic qubit and the ac-Stark effect of spin qubit}

According to Eqs.~({\color{blue}9}) and ({\color{blue}10}), we rewrite Hamiltonian ({\color{blue}8}) as
\begin{equation}
\begin{aligned}
\hat{H}\approx&\hat{H}'_0-\hbar\Omega_{12}\left(|1\rangle\langle 2|+{\rm h.c.}\right)
\cos(\omega_T t-k_Tx)\\
&
+\hbar\eta_0\omega_s(\hat{a}_x^\dagger+\hat{a}_x)\hat{S}_x\,,
\end{aligned}
\end{equation}
with the revised free-Hamiltonian
\begin{equation}
\begin{aligned}
\hat{H}'_0=&\hat{H}_0-\frac{\hbar\omega_s}{2}\hat{S}_z
+\hbar\omega_s(\eta_{1}|1\rangle\langle 1|+\eta_{2}|2\rangle\langle 2|)\hat{S}_z\,,
\end{aligned}
\end{equation}
and where $\eta_0=x_0/(2h_0)\ll1$, $\eta_{1}=z_{11}/(2h_0)\ll1$, and $\eta_{2}=z_{22}/(2h_0)\ll1$.

In the interaction picture defined by
$\exp(-i\hat{H}'_0t/\hbar)$, Hamiltonian ({\color{blue}11}) can be further written as
\begin{equation}
\begin{aligned}
\hat{H}_{\rm int}\approx&\hbar\Omega_{12}\left[e^{i(\Delta_a-\Delta_s)t}|1\rangle\langle 2|+{\rm h.c.}\right]\otimes|\downarrow\rangle\langle \downarrow|\\
&+\hbar\Omega_{12}\left[e^{i(\Delta_a+\Delta_s)t}|1\rangle\langle 2|+{\rm h.c.}\right]\otimes|\uparrow\rangle\langle \uparrow|\\
&-\hbar\eta_0\omega_s\left[\hat{a}_x^\dagger|\downarrow\rangle\langle\uparrow|
e^{i(\Delta_{x}+2\eta_{1}\omega_s)t}
+{\rm h.c.}\right]
\otimes|1\rangle\langle1|\\
&-\hbar\eta_0\omega_s\left[\hat{a}_x^\dagger|\downarrow\rangle\langle\uparrow|
e^{i(\Delta_{x}+2\eta_{2}\omega_s)t}
+{\rm h.c.}\right]
\otimes|2\rangle\langle2|\,,
\end{aligned}
\end{equation}
under the well-known rotating wave approximation and the Lamb-Dicke approximation $\Omega_{12}\exp(-ik_Tx)\approx\Omega_{12}$~\cite{LD}.
In Hamiltonian ({\color{blue}13}), $\Delta_a=\omega_T-\omega_a$ denotes the detuning between the inputting mm wave and the hydrogen-like atom, $\Delta_s=\omega_{s}(\eta_2-\eta_1)$ stands for the spin-dependent Stark shift of such an atom, and $\Delta_{x}=\omega_{x}-\omega_s$ describes the detuning between spin and the lateral harmonic oscillator. Numerically, using the aforementioned $h_0=5\,\,\mu{\rm m}$, $\omega_s\approx3.5$~GHz, and $\omega_x\approx12$~GHz, we have $\Delta_s\approx8.4$~MHz and $\Delta_{x}\approx8.5$~GHz, with  $\eta_2-\eta_1=(z_{22}-z_{11})/(2h_0)\approx2.4\times10^{-3}$ and $z_{22}-z_{11}\approx3.2\,z_b$. Given the adjustable $\Delta_a$, $\Delta_s$, and $\Delta_{x}$ with the controllable bias-votage and bias-current, we highlight the following two solutions.

Firstly, for $\Delta_a-\Delta_s=0$, $\Delta_s\gg\Omega_{12}$, and $\omega_x\gg\omega_s$, Hamiltonian ({\color{blue}13}) can be reduced to
\begin{equation}
\begin{aligned}
\hat{H}_{\rm d}=
\hbar\left(\Omega_{12}|1\rangle\langle 2|+{\rm h.c.}\right)\otimes|\downarrow\rangle\langle\downarrow|\,,
\end{aligned}
\end{equation}
by using the rotating wave approximation again.
This Hamiltonian results in the oscillating height of electron above the liquid helium surface and then generates the experimentally detectable ac currents~\cite{DenisPRL2019,DenisJLTP2021}. However, Hamiltonian ({\color{blue}14}) is not used as the mechanism of mm wave sensing, due to the strong decay of the excited orbital
state $|2\rangle$~\cite{Saturation}. Instead, Hamiltonian ({\color{blue}14}) with a sufficiently large Rabi frequency $\Omega_{12}$ will be used to detect the spin state $|\downarrow\rangle$.

Secondly, for the desired mm wave precise detection, we set the system working within the large-detuning regime, i.e., $\Delta_a\pm\Delta_s\gg\Omega_{12}$ and $\omega_x\gg\omega_s$. In such a regime, Hamiltonian ({\color{blue}13}) can be approximated written as
\begin{equation}
\begin{aligned}
\hat{H}_{\rm int}\approx&\hbar\Omega_{sz\downarrow}|\downarrow\rangle\langle\downarrow|
\otimes
\left(|1\rangle\langle1|-|2\rangle\langle2|\right)\\
&
+
\hbar\Omega_{sz\uparrow}|\uparrow\rangle\langle\uparrow|
\otimes
\left(|1\rangle\langle1|-|2\rangle\langle2|\right)
\\
&
-\hbar\Omega_{sx1}\hat{a}_x^\dagger\hat{a}_x
(|\uparrow\rangle\langle\uparrow|-|\downarrow\rangle\langle\downarrow|)
\otimes|1\rangle\langle1|\\
&-\hbar\Omega_{sx2}\hat{a}_x^\dagger\hat{a}_x
(|\uparrow\rangle\langle\uparrow|-|\downarrow\rangle\langle\downarrow|)
\otimes|2\rangle\langle2|
\\
&-\hbar\Omega_{sx1}|\uparrow\rangle\langle\uparrow|
\otimes|1\rangle\langle1|
\\
&-\hbar\Omega_{sx2}|\uparrow\rangle\langle\uparrow|
\otimes|2\rangle\langle2|\,,
\end{aligned}
\end{equation}
with $\Omega_{sz\downarrow}=\Omega_{\rm 12}^2/(\Delta_a-\Delta_s)$, $\Omega_{sz\uparrow}=\Omega_{\rm 12}^2/(\Delta_a+\Delta_s)$, $\Omega_{sx1}=\eta_0^2\omega_s^2/[\omega_x-\omega_s(1-2\eta_1)]$, and
$\Omega_{sx2}=\eta_0^2\omega_s^2/[\omega_x-\omega_s(1-2\eta_2)]$.

The obtained Hamiltonian ({\color{blue}15}) does not excites any orbital states. In the low temperature about $10$~mK~\cite{SchusterNature}, both lateral harmonic oscillator (about $12$~GHz) and the normal hydrogen-like atom can be cooled well into their ground states. Therefore, Hamiltonian ({\color{blue}15}) reduces greatly to
\begin{equation}
\begin{aligned}
\hat{H}_{\rm int}=&\hbar\Omega_{sz\downarrow}
|\downarrow\rangle\langle\downarrow|
+\hbar(\Omega_{sz\uparrow}-\Omega_{sx1})|\uparrow\rangle\langle\uparrow|\,.
\end{aligned}
\end{equation}
Note that, $\Omega_{sx1}$ refers to the zero-point fluctuation of the orbital harmonic oscillation, which results only in a small renormalization of the transition frequency of the spin qubit. In the interaction picture defined by $\exp(it\Omega_{sx1}|\uparrow\rangle\langle\uparrow|)$, Hamiltonian ({\color{blue}16}) becomes
\begin{equation}
\begin{aligned}
\hat{H}_{\rm eff}=-\frac{\hbar\Omega_{sz}}{2}\hat{S}_z\,,
\end{aligned}
\end{equation}
with
\begin{equation}
\begin{aligned}
\Omega_{sz}&=\Omega_{sz\uparrow}-\Omega_{sz\downarrow}=\frac{2\Omega_{12}^2\Delta_s}{\Delta_a^2-\Delta_s^2}\,.
\end{aligned}
\end{equation}
This is nothing but the ac-Stark effect of spin generated by the electric dipole interaction between the mm wave and the 1D hydrogen-like atom.

\section{Hahn echo interferometry for mm wave sensitive detection}
Now, let us discuss how to implement the sensitive detection of mm wave
electric field by using the above spin ac-Stark effect ({\color{blue}17}) and the Hahn echo interferometry~\cite{Hahn-echo1,Hahn-echo2} as well as the detection method for surface electrons~\cite{DenisPRL2019}.
The detecting sensitivity of this proposal is consequently analyzed by considering the well-known Johnson-Nyquist white noise and the spin projective
measurement noise.

\subsection{The phase shift due to spin ac-Stark effect}
\begin{figure}
\subfigure[]{\includegraphics[width=7cm]{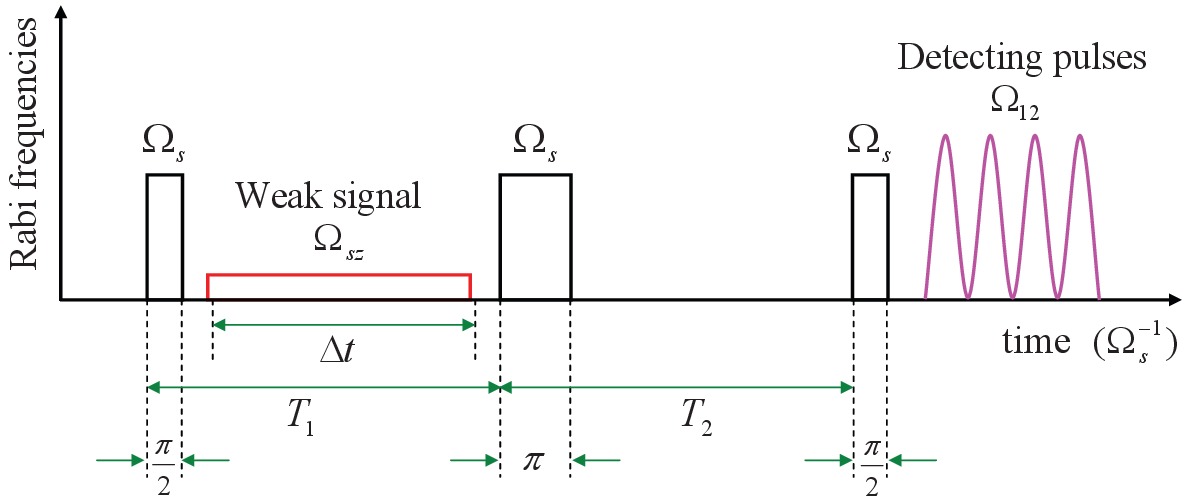}}
\subfigure[]{\includegraphics[width=7cm]{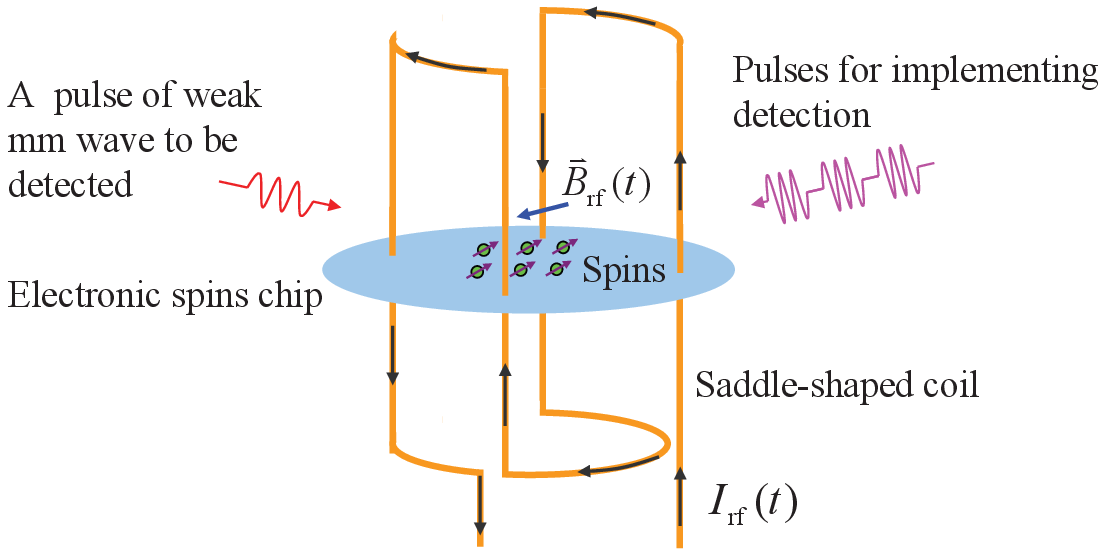}}
\caption{(a) Operational sequence of spin qubit for detecting the signal $\Omega_{sz}\Delta t$ of a weak mm wave. After $\pi/2-\Omega_{sz}\Delta t-\pi-\pi/2$ sequence, signal $\Omega_{sz}\Delta t$ has been stored into the spin superposition state. To detect such signal, the additional pulses of strong mm wave are applied to periodically excite electronic orbital state. Such excitation is spin dependent, and then the signal $\Omega_{sz}\Delta t$ is experimentally detectable~\cite{DenisPRL2019}. (b) The saddle-shaped coil~\cite{Saddle-coil} is suitable to implement the desired $\pi/2$ and $\pi$ operations for the present spins chip, as the RF magnetic field $\textbf{B}_{\rm rf}$ generated by this device has high uniformity and can also be perpendicular to the static magnetic field showed in Sec. III.}
\label{fig1}
\end{figure}
Firstly, we use the standard Hahn echo sequence $\pi/2-\Omega_{sz}\Delta t-\pi-\pi/2$~\cite{NV-RMP} to store the signal $\Omega_{sz}\Delta t$, as shown schematically in Fig.~\textcolor{blue}{3(a)}. The mm wave signal is inserted between the first $\pi/2$ and the middle $\pi$ pulses.
Basically, the single-qubit operations of the spin qubit, i.e., the so-called $\pi/2$ and $\pi$ pulses, are the key steps in the spin interferometry. Given the magnetic resonance imaging (MRI) technology has been widely used in the medical field, we also use the oscillating magnetic fields to implement the desired single-qubit operations, via the mechanism of electron spin resonance (ESR)~\cite{Hahn-echo1,Hahn-echo2}. In the field of MRI, many kinds of Radio-frequency (RF) coils such as the famous birdcage-shaped coils and the saddle-shaped coils, were designed to generate the oscillating magnetic fields with high spatial-uniformity for implementing the high-fidelity medical image~\cite{MRI}.

As a simple case showed in Fig.~\textcolor{blue}{3(b)}, the saddle-shaped coil~\cite{Saddle-coil} can be used to generate the desired RF magnetic field for implementing the ESR. Supposing the transition frequency of the spin qubit is set at $\omega_s\approx0.4$~GHz by adiabatically controlling the
bias-current $I_{\rm dc}$, then the wavelength of the required RF fields are far larger than the
distribution range of the trapped electrons, and thus the resonant RF magnetic field $\textbf{B}_{\rm rf}(t)$ near the center of the coil can be treated as the spatially uniform one, i.e.,
$\textbf{B}_{\rm rf}(t)
\approx B_y\sin(\omega_st+\vartheta)\textbf{e}_y$, with $B_y$ and $\vartheta$ being its amplitude and initial phase, respectively. As a result, the Hamiltonian describing ESR can be written as
\begin{equation}
\hat{H}_{\rm ESR}
\approx\frac{\hbar\Omega_s}{2}\left(e^{i\vartheta}|\downarrow\rangle\langle \uparrow|
+{\rm h.c.}\right)
\end{equation}
in the interaction picture. Here, $\Omega_s=u_bB_{y}/\hbar$ is the spin Rabi frequency, and numerically, $\Omega_s\approx1$~MHz, with $B_y=11\,\mu {\rm T}$ and $u_b\approx9.3\times10^{-24}\,{\rm A}{\rm m}^2$ being the Bohr magneton.

In Schr\"{o}dinger picture, the spin dynamics with Hamiltonian ({\color{blue}19}) can be described by the time evolution operator $\hat{U}_s=\hat{U}_0(t)\hat{R}(\Omega_st)$. Here, $\hat{U}_0(t)=\exp(-i\omega_{rs} t\hat{S}_z/2)$ is the free-evolution operator of spin qubit, with $\omega_{rs}=\omega_s-2\eta_1\omega_s-\Omega_{sx1}$ being the renormalized transition frequency of the qubit dressed by the orbital ground states.
The evolution operator $\hat{R}(\Omega_st)=\exp(-i\Omega_st\hat{S}_y/2)$, with $\hat{S}_y=\exp(i\vartheta)|\downarrow\rangle\langle \uparrow|
+{\rm h.c.}$, is used to either generate the superposition or flip the spin qubit. It is worth to note that the additional phase $\vartheta$ is eliminable by the following Hahn echo operation.

The time evolution due to the operational sequence showed by Fig.~\textcolor{blue}{3(a)} reads
\begin{equation}
\begin{aligned}
\hat{U}_{\rm {echo
}}=&\hat{R}(\pi/2)\hat{U}_0(T_2)\hat{R}(\pi)\hat{U}_0(T_1)
\hat{U}_{sz}(\Delta t)\hat{R}(\pi/2)\,,
\end{aligned}
\end{equation}
with $\hat{U}_{sz}(\Delta t)=\exp(i\hat{S}_z\Omega_{sz}\Delta t/2)$
being the evolution due to the mm wave inserted between the first $\pi/2$ pulse and the middle $\pi$ pulse. $T_1$ and $T_2$ are the durations of the two-stage free evolutions. The duration of the inserted mm wave driving is set to satisfy the condition $T_1>\Delta t\gg\delta t$, with $\delta t=\pi/(2\Omega_s)$ being the duration of the so-called $\pi/2$ pulse.

Consider that the spin qubit is initially prepared at its ground
state, then the final state after the Hahn echo sequence reads $|f\rangle=\hat{U}_{\rm {echo
}}|\downarrow\rangle$, wherein the probability of finding the spin ground state reads
\begin{equation}
\begin{aligned}
P=|\langle \downarrow|f\rangle|^2=\frac{1}{2}
+\frac{1}{2}\cos(\Omega_{sz}\Delta t+\theta_0)\,,
\end{aligned}
\end{equation}
with $\theta_0=(T_2-T_1)\omega_{rs}$.
In order to measure the probability $P$ (and thus the phase shift $\Omega_{sz}\Delta t$ due to the signal wave), another mm wave with a different carrier-frequency is expected to excite the orbital state of trapped electron. The excitation is spin $|\downarrow\rangle$ dependent, as shown by Hamiltonian ({\color{blue}14}), and then the
signal $\Omega_{sz}\Delta t$ is experimentally detectable. We explain this below, in detail.

\subsection{The spins detection with trapped electrons on liquid helium}
In general, the vibration of a trapped charge is dissipative, due to the interaction with surroundings. The circuit with resistance $R$ showed by Fig.~\textcolor{blue}{1(a)} is a surrounding with the dissipated power $P_R=i(t)u(t)$, which generates observable voltage $u(t)=i(t)R$ with the current~\cite{Penning,Leibfried2022}
\begin{equation}
\begin{aligned}
i(t)=\frac{\kappa e}{l}\frac{{\rm d}z(t)}{{\rm d}t}\,.
\end{aligned}
\end{equation}
Here, $l$ is the width of the micro-box showed by Fig.~\textcolor{blue}{1(a)}, and $\kappa$ is a parameter used to describe $z$-directional electric field $\kappa u(t)/l$ which damps the vibrational height $z(t)$ of electron on liquid helium. According to the numerical computation in Sec. II, we find $\kappa/l\approx0.22\times10^{5}\,{\rm m}^{-1}$.

The above equation is a classical presentation applied earlier for describing the signal of trapped charges in the Penning trap~\cite{Penning}. For the quantum counterpart of Eq. {\color{blue}(22)}, we replace $i(t)$ and $z(t)$ by $\hat{i}(t)$ and $\hat{z}(t)$, respectively. Correspondingly, Eq.~{\color{blue}(22)} gives rise to the following solution,
\begin{equation}
\begin{aligned}
\hat{i}(t)&=\frac{\kappa e}{l}\frac{{\rm d}}{{\rm d}t}\left[\frac{1}{2\pi}\int_{-\infty}^{\infty}\hat{z}(\omega)e^{i\omega t}{\rm d}\omega\right]\,,
\end{aligned}
\end{equation}
with the Fourier transform
\begin{equation}
\begin{aligned}
\hat{z}(\omega)
=\int_{-\infty}^{\infty} \hat{z}(t) e^{-i\omega t}{\rm d}t
=\int_{0}^{\infty} \hat{z}(t) e^{-i\omega t}{\rm d}t\,,
\end{aligned}
\end{equation}
and the single side signal $\hat{z}(t)\neq{\rm constant}$ in time domain $t>0$.

More clearly, we rewrite Eq.~{\color{blue}(23)} as
\begin{equation}
\begin{aligned}
\hat{i}(t)=\int_{0}^{\infty}\left[\hat{j}(\omega) \cos(\omega t)-\hat{i}(\omega)\sin(\omega t)\right]{\rm d}
\omega\,,
\end{aligned}
\end{equation}
with the quantized currents
\begin{equation}
\begin{aligned}
&\hat{j}(\omega)
=\frac{\kappa e\omega}{\pi l}\int_{0}^{\infty} \hat{z}(t)\sin(\omega t){\rm d}t\,,
\end{aligned}
\end{equation}
and
\begin{equation}
\begin{aligned}
\hat{i}(\omega)=
\frac{\kappa e\omega}{\pi l}\int_{0}^{\infty} \hat{z}(t)\cos(\omega t){\rm d}t\,.
\end{aligned}
\end{equation}
In principle, the lock-in amplifier can distinguish the phase difference between $\cos(\omega t)$ and $\sin(\omega t)$. This means that the signals
$\hat{j}(\omega)$ and $\hat{i}(\omega)$ can be measured independently. In what follows, we specifically treat $\hat{i}(\omega)$, as it is the same as for $\hat{j}(\omega)$.

As usual, the trapped electron should be driven to produce the persistent signals. Specially,
Hamiltonian ({\color{blue}14}) indicates that the orbital driving is related to the spin state $|\downarrow\rangle$, thus
Eq.~({\color{blue}27}) becomes
\begin{equation}
\begin{aligned}
\hat{i}_s(\omega)&=|\downarrow\rangle\langle \downarrow|\otimes\hat{i}(\omega)\,,
\end{aligned}
\end{equation}
and then its expectant value reads
\begin{equation}
\begin{aligned}
\langle \hat{i}_s(\omega)\rangle
&=P\frac{\kappa e\omega}{\pi l}\int_{-\infty}^{\infty} \langle\hat{z}(t)\rangle \cos(\omega t){\rm d}t\\
&=P\frac{\kappa e\omega}{\pi l}\int_{-\infty}^{\infty} \text{Tr}[\hat{z}\hat{\rho}(t)]\cos(\omega t){\rm d}t\\
&\approx P\frac{\kappa e\omega}{\pi l}(z_{22}-z_{11})\int_{-\infty}^{\infty} \rho_{22}(t) \cos(\omega t){\rm d}t\\
&=Pq_0\omega\rho_{22}(\omega)\,.
\end{aligned}
\end{equation}
In the second line of this equation, $\hat{\rho}(t)=\sum_{i,j}^2\rho_{ij}(t)|i\rangle\langle j|$ is the density operator of the orbital qubit of electron floating on liquid helium. The matrix elements $\rho_{ij}(t)$ satisfy the well-known relations $\sum_{i=1}^2\rho_{ii}(t)=1$ and $\rho_{ij}(t)=\rho_{ji}^{\ast}(t)$. In the third line of Eq.~({\color{blue}29}), $z_{22}-z_{11}=\langle 2|\hat{z}|2\rangle-\langle 1|\hat{z}|1\rangle$ is the relative height between the excited state electron and the ground state one. The terms related to the off-diagonal elements $z_{12}=\langle 1|\hat{z}|2\rangle$ and $z_{21}=\langle 1|\hat{z}|2\rangle$ are quickly oscillating and have been neglected for the detection of low frequency signals~\cite{DenisPRL2019,DenisJLTP2021}. In the last line of Eq. ({\color{blue}29}), $q_0=\kappa e(z_{22}-z_{11})/l$ is an effective charge.

It is noteworthy that, Eq.~({\color{blue}29}) indicates that the unit of $\langle \hat{i}_s(\omega)\rangle$ is ampere per Hertz. Nevertheless, it can be regarded as an observable current by using the dimensionless integral
\begin{equation}
\begin{aligned}
\rho_{22}(\omega)&=\frac{{\rm d}\omega}{\pi}\int_{-\infty}^{\infty} \rho_{22}(t) \cos(\omega t){\rm d}t\\
&=\frac{1}{T}\int_{-T}^{T} \rho_{22}(t) \cos(\omega t){\rm d}t\,,
\end{aligned}
\end{equation}
with ${\rm d}\omega/\pi=1/T$ and $T\rightarrow\infty$. This form of Fourier transform can be easily checked by inserting the test function $\rho_{22}(t)=\cos(\omega t)$ into the integral. It gives the expected $\rho_{22}(\omega)\rightarrow1$.

Considering now $N$ electrons, Eq.~{\color{blue}(28)} becomes
\begin{equation}
\begin{aligned}
\hat{I}(\omega)=\sum_{n=1}^{N}|\downarrow_n\rangle\langle \downarrow_n|\otimes\hat{i}_n(\omega)\,,
\end{aligned}
\end{equation}
and then the signal amplitude reads
\begin{equation}
\begin{aligned}
I(\omega)=\sum_{n=1}^{N}P_n\langle \hat{i}_n(\omega)\rangle\approx PNq_0\omega \rho_{22}(\omega)\,.
\end{aligned}
\end{equation}
This formula is similar to that demonstrated by the experiment~\cite{DenisPRL2019}, except the probability $P$ of spin state $|\downarrow\rangle$ considered in this work.

\begin{figure}[tbp]
\includegraphics[width=7cm]{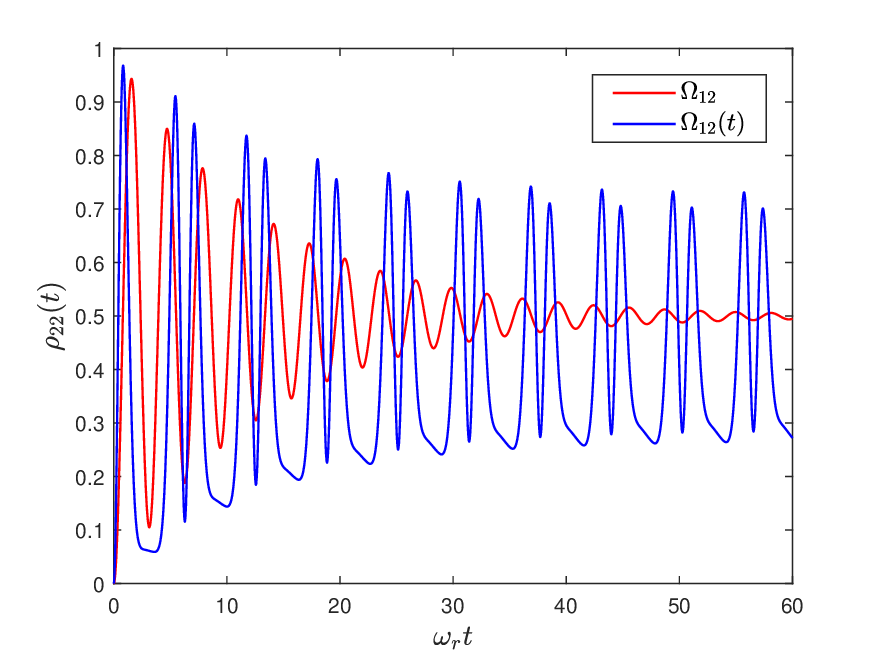}
\caption{Numerical solutions of $\rho_{22}(t)$. The red and the blue lines show respectively that the constant Rabi frequency $\Omega_{12}$ can not generate the persistently oscillating $\rho_{22}(t)$, but the time-dependent Rabi frequency $\Omega_{12}(t)=[1+\cos(\omega_r t)]\Omega_{12}$ can. To plot this figure, we assigned the modulation frequency $\omega_{r}=\Omega_{12}$ and the orbital decay frequency $\Gamma_a=0.1\Omega_{12}$.}
\end{figure}
Consequently, we solve the well-known Lindblad master equation
\begin{equation}
\begin{aligned}
\frac{{\rm d}\hat{\rho}}{{\rm d}t}=&\frac{-i}{\hbar}[\hat{H}_{\rm d},\hat{\rho}]
+\frac{\Gamma_a}{2}(2\rho_{22}|1\rangle\langle1|-|2\rangle\langle2|\hat{\rho}
-\hat{\rho}|2\rangle\langle2|)\,
\end{aligned}
\end{equation}
to get the density matrix element $\rho_{22}(t)$ and its Fourier transform $\rho_{22}(\omega)$. Note that, $\hat{H}_{\rm d}$ has been given by Eq.~({\color{blue}14}), and $\Gamma_a$ is a phenomenological decay rate of atomic exciting state~\cite{Wang}.
The master equation yields
\begin{equation}
\begin{aligned}
&\frac{{\rm d}\rho_{22}}{{\rm d}t}
=\left(i\Omega_{12}\rho_{21}+{\rm c.c.}\right)
-\Gamma_a\rho_{22}\,,\\
&\frac{{\rm d}\rho_{21}}{{\rm d}t}=i\Omega_{12}(2\rho_{22}-1)
-\frac{\Gamma_a}{2}\rho_{21}\,.
\end{aligned}
\end{equation}

For an arbitrary Rabi frequency $\Omega_{12}$, the above equations
can be solved by the numerical method, and consequently, $\rho_{22}(\omega)$ can be computed by the following numerical integral,
\begin{equation}
\begin{aligned}
\rho_{22}(\omega)
&=\frac{1}{T}
\int_{0}^{T}\rho_{22}(t)\cos(\omega t) {\rm d}t
\\
&=\frac{1}{N}\sum_{n=0}^{N}\rho_{22}(n{\rm d}t)\cos(\omega n{\rm d}t)\,,
\end{aligned}
\end{equation}
with $T=N{\rm d}t\rightarrow\infty$, $N\rightarrow\infty$, and ${\rm d}t\rightarrow0$.

\begin{figure}[tbp]
\includegraphics[width=7cm]{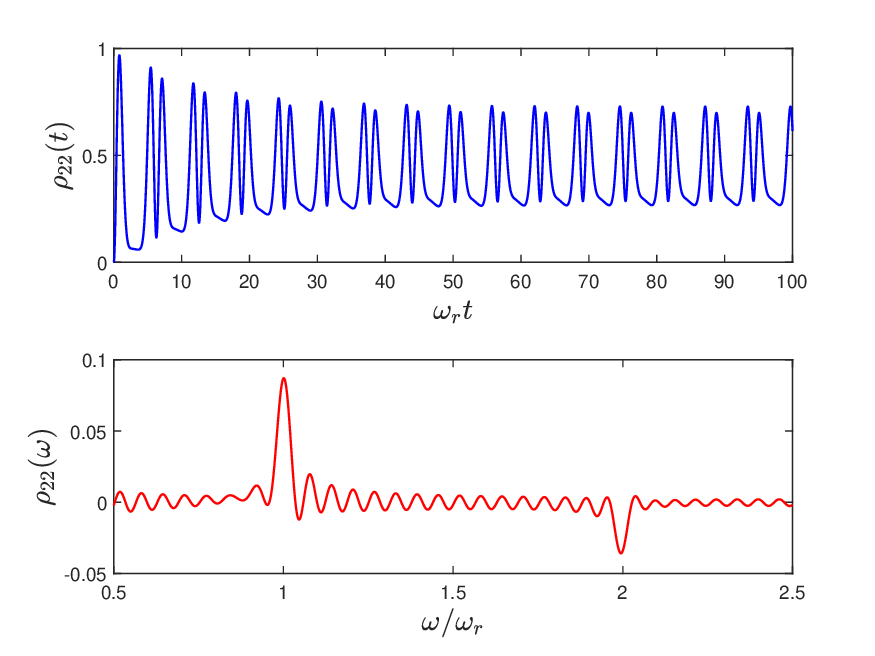}
\caption{Density matrix element $\rho_{22}$ is plotted within time and frequency domains, with the time-dependent Rabi frequency $\Omega_{12}(t)=[1+\cos(\omega_r t)]\Omega_{12}$. The essential values of $\Omega_{12}$ and $\Gamma_a$ for this figure are the same as that used in Fig.~\textcolor{blue}{4}. The solution shows a significant signal which oscillates with the modulation frequency $\omega_r$ of the pumping wave.}
\end{figure}

For a constant Rabi frequency $\Omega_{12}$, Fig.~\textcolor{blue}{4} shows that the $\rho_{22}(t)$-oscillation (and thus the observable signal) is damped quickly, due to the unavoidable decay $\Gamma_a$.
To produce the persistent ac currents, an amplitude modulated pumping wave is necessary~\cite{DenisPRL2019}. This means the effective Rabi frequency being time-dependent specially such as $\Omega_{12}\rightarrow\Omega_{12}(t)=[1+\cos(\omega_r t)]\Omega_{12}$. For this modified Rabi frequency, Fig.~\textcolor{blue}{5} shows clearly an ac current which oscillates with the
modulation frequency $\omega_r$. This result is consistent with the experimental phenomenon reported in Ref.~\cite{DenisPRL2019}.

\subsection{The sensitivity of mm wave electric field detection}
It is emphasized that, as the Rabi frequency $\Omega_{12}$ is required to be comparable to the orbital decay rate, the above orbital excitation of surface-state electron is suitable to detect the spin state, instead of the very weak mm wave directly. Given the coherent time of the spin qubit is generically longer than that of atomic qubit, we use the spin ac-Stark effect ({\color{blue}17}) to measure the weak mm wave. The possible sensitivity of this measurement is estimated by using the following approach.

Firstly, Eq.~({\color{blue}21}) for the single spin should be corrected as
\begin{equation}
\begin{aligned}
P=\frac{1}{2}+\frac{1}{2}e^{-\Gamma t}\cos(\Omega_{sz} t+\theta_0)\,,
\end{aligned}
\end{equation}
due to the practical existing interference damping, i.e., $\exp(-\Gamma t)$. Here, $\Gamma$ stands for the damping speed of interference, and $1/\Gamma$ refers to the coherence time. As we well-known, the decoherence of a qubit is caused by the so-called decay and the dephasing. The former refers to the energy transition between qubit and its surroundings. The latter is due to the surroundings-disturbed phase in qubit's superposition state.
For the spin qubit of trapped electron on liquid helium, the dephasing is the major factor resulting in decoherence~\cite{Lyon,Schuster2010}.

Secondly, supposing the mm wave is very weak, i.e., $\Omega_{sz}t\ll1$, then Eq.~({\color{blue}36}) reduces to
\begin{equation}
\begin{aligned}
P\approx\frac{1}{2}+\frac{1}{2}e^{-\Gamma t }
\Omega_{sz}t\,,
\end{aligned}
\end{equation}
within the so-called slope regimes around $\theta_0\approx\pi/2,3\pi/2,\cdots$~\cite{Quantum sensing}.
According to Eq.~({\color{blue}32}), the ac-Stark effects of $N$ spins deliver the signal
\begin{equation}
\begin{aligned}
I_s(\omega)\approx\frac{1}{2}e^{-\Gamma t }\Omega_{sz}t I_0(\omega)\,,
\end{aligned}
\end{equation}
with
\begin{equation}
\begin{aligned}
I_0(\omega)=Nq_0\omega \rho_{22}(\omega)\,.
\end{aligned}
\end{equation}
As an experimentally detectable quantity, signal ({\color{blue}38}) should be comparable to the noise current during the stage of detection. Using $\tilde{I}(\omega)$ to denote such a noise and letting $I_s(\omega)=\tilde{I}(\omega)$, then
the minimum detectable effect of ac-Stark shift is gotten as
\begin{equation}
\begin{aligned}
\left[\Omega_{sz}\right]_{\text{min}}
\approx\left(\frac{2 e^{\Gamma t }}{t}\right)_{\text{min}}
\frac{\tilde{I}(\omega)}{I_0(\omega)}=\frac{2\mathcal{N}}{t_c}\,.
\end{aligned}
\end{equation}
Here, we have used the mathematical relation $[t^{-1}\exp(\Gamma t)]_{\text{min}}={\rm e}\Gamma$, with ${\rm e}\approx2.7$ being the Euler number. As a consequence, $t_c=1/({\rm e}\Gamma)$ refers to the coherent time of spin qubit. Furthermore, we have $\tilde{I}(\omega)/I_0(\omega)=\mathcal{N}$ with $1/\mathcal{N}$ being the signal-noise ratio.

Thirdly, according to Eq.~({\color{blue}18}) of ac-Stark shift, the minimum detectable strength of mm wave electric field is described by
\begin{equation}
\begin{aligned}
E_{\text{min}}
\approx\frac{\hbar}{\mu_{12}}\sqrt{\frac{(\Delta_a^2-\Delta_s^2)\mathcal{N}}{\Delta_s t_c}}\,,
\end{aligned}
\end{equation}
with $\mu_{12}=ez_{12}\approx0.5ez_b\approx6.0\times10^{-28}\,{\rm C\,m}$~\cite{Dykman-qubit} being the electric dipole moment of electronic orbital motion. Suppose that the frequency of the mm wave is set as $\Delta_a=\Delta_s+\delta_{as}$ with $\delta_{as}\ll\Delta_s$, then Eq.~({\color{blue}41}) reduces to
\begin{equation}
\begin{aligned}
\left(\frac{E}{\sqrt{\delta_{as}}}\right)_{\text{min}}
\approx\frac{\hbar}{\mu_{12}}\sqrt{\frac{2\mathcal{N}}{t_c}}\,.
\end{aligned}
\end{equation}
Mathematically, we can write $\delta_{as}=\xi\Omega_{12}$ with a parameter $\xi$, and therefore, the above formula becomes
\begin{equation}
\begin{aligned}
E_{\text{min}}
\approx\frac{\hbar}{\mu_{12}}\frac{2\xi\mathcal{N}}{t_c}\,.
\end{aligned}
\end{equation}
Physically, parameter $\xi$ refers to the orbital excitation due to the driving of mm wave, i.e., the validity of the rotating wave approximation made in Sec. III-B. Typically, $\xi=10$ is large enough to suppress orbital excitation as the probability of excited orbital state is at the order of $1/\xi^2$ (which results in a negligible correct to the total number of the available qubits).
This allows us to get the above sensitivity without the orbital coherent time which is generically shorter than the spin coherent times such as $t_c\sim1\,\,\text{ms}$~\cite{Schuster2010}, see Appendix {\color{blue} B}, in detail.

Finally, to calculate the noise-signal ratio $\mathcal{N}$, we write $\tilde{I}^2(\omega)=I_J^2+I_o^2(\omega)$ in terms of two typical noise-currents: the Johnson-Nyquist white noise $I^2_J$, and the shot noise $I^2_o(\omega)$. Therefore, the noise-signal ratio reads
\begin{equation}
\begin{aligned}
\mathcal{N}=\sqrt{\frac{\tilde{I}^2(\omega)}{I_0^2(\omega)}}=\sqrt{\frac{4k_bT\Delta\omega}{RI_0^2(\omega)}+\frac{I_o^2(\omega)}{I_0^2(\omega)}}\,.
\end{aligned}
\end{equation}
As we well-known, Johnson-Nyquist current noise reads $I_J^2=4k_bT\Delta\omega/R$, with $T$ being the temperature of the resistance $R$, and $\Delta \omega$ the measurement bandwidth (around the interested frequency $\omega$ of the output signal). Johnson-Nyquist noise is a white noise, as its power spectral density $4k_bT$ is independent on the frequency. Actually, the term $4k_bT\Delta\omega/[RI_0^2(\omega)]$ in Eq.~({\color{blue}44}) is the ratio between Johnson-Nyquist power and the signal power. This means that whether treating current or voltage as the signal, the noise-signal ratio $\mathcal{N}$ is unchanged.

On the other hand, the so-called shot noise $I_o^2(\omega)$ in Eq.~({\color{blue}44}) is originated from the spin projective measurement~\cite{Wineland-SQL,Science-SQL,JPA-SQL} as well as the orbital quantum motions of the driving surface-state electrons on liquid helium. In detail, the shot noise is defined by the variance $I_o^2(\omega)=\langle\hat{I}^2(\omega)\rangle-\langle\hat{I}(\omega)\rangle^2$. According to
Eq.~({\color{blue}31}), we have
\begin{equation}
\begin{aligned}
I_o^2(\omega)&=\sum P_n\langle\hat{i}_n^2(\omega)\rangle-P_n^2\langle\hat{i}_n(\omega)\rangle^2\\
&\approx N P\langle\hat{i}^2(\omega)\rangle-NP^2\langle\hat{i}(\omega)\rangle^2\\
&\approx N(P-P^2)\langle\hat{i}(\omega)\rangle^2\,.
\end{aligned}
\end{equation}
In the second line of this equation, we have used the identical-particles approximation. For the third line, we have used a rough relation $\langle\hat{i}^2(\omega)\rangle\approx\langle\hat{i}(\omega)\rangle^2$ based on Appendix~\textcolor{blue}{C} which refers to the two-time correlation function of the induced currents.

Given $I_o^2(\omega)<N\langle\hat{i}(\omega)\rangle^2$ in Eq.~({\color{blue}45}),
we would like to reexpress Eq.~({\color{blue}44}) simply as
\begin{equation}
\begin{aligned}
\mathcal{N}=\sqrt{\frac{4k_bT\Delta\omega}{RI_0^2(\omega)}+\frac{1}{N}}\,.
\end{aligned}
\end{equation}
Numerically, we have $I_0(\omega)=Nq_0\omega \rho_{22}(\omega)\approx0.68\,{\rm pA}$, with $\omega=1$~MHz, $q_0=5.4\times10^{-4}\,e$, $\rho_{22}(\omega)=0.1$, and $N=8\times10^4$. As a consequence, $4k_bT\Delta\omega/[RI_0^2(\omega)]\approx1.2\times10^{-4}$ and $\mathcal{N} \approx0.01$, with $\Delta\omega=1\,{\rm Hz}$, $R=10\,{\rm k\Omega}$, and $T=10$~mK. Using these values and that listed blow Eqs.~({\color{blue}41}) and~({\color{blue}43}), we have $E_{\text{min}}
\approx39\,\,\mu\text{Volt}/\text{m}$, which reaches the level of using Rydberg atoms as the mm waves sensor~\cite{Rydberg-mm}.

Certainly, the less bandwidth $\Delta\omega$ refers to the lower Johnson-Nyquist power. However, a more smaller bandwidth requires more longer detection time for successful spectrum analysis. On the other hand, Eq.~({\color{blue}46}) tells us that the large resistance $R$ is useful to enhance the output power~\cite{Penning}. However, the large output power leads also to the strong dissipation of the electronic orbital motion, for example, $\Gamma_a=(\kappa e/l)^2R/m_e\approx0.14\,{\rm MHz}$ with $R=10\,{\rm k\Omega}$, which is significantly larger than that due to the spontaneous radiative emission~\cite{Science-qubit,Dykman-qubit}.
A more larger resistance requires also the stronger detection field for enhancing the stimulated ac signals.
However, the driving electric field (i.e., the Rabi frequency in detecting stage) can not be too large, due to limited strength of spin-orbit coupling used for spin projective measurement.
Enlarging $N$ may be a direct approach to improve the detecting sensitivity. However, this requires the more large sizes of the chip, and then is limited by the effective area of mm wave illuminating as well as the non-uniformity of ESR.

Given the above limitations, optimizable coherent time of spin could be an advantage of the present system. As it shown in~\cite{Schuster2010} and also in Appendix {\color{blue} B}, the fluctuating height of helium surface with respect to the basis current is the main noise source that leads to spin decoherence. This means that while increasing the spin-orbit coupling, the applied current increases also the decoherence. Let us see the derivation for Eq.~({\color{blue}43}), wherein the strength of spin-orbit coupling is not required so large as that for spin projective measurement. This means that, to improve the detecting sensitivity, one can use a relatively weak spin-orbit coupling during the stage of Hahn echo sequence, and consequently enhance the spin-orbit coupling for implementing the efficient detection for spins.
The operations would be complex in practical experiments, as the adjustable bias-current is required. We hope this is not a fundamental problem.

\section{Conclusion}
In summary, we have introduced a conceptually new approach to implement the sensitive detection of mm wave electric field by using the spin ac-Stark effect of trapped electrons on liquid helium. To achieve the proposal, we designed a circuit chip to trap multiple electrons individually on superfluid helium film, and then, a dc-current is loaded very close to every electron for generating the spin qubit and also the strong spin-orbit coupling for each of the trapped electrons.
The proposal is based on the fact that the transition frequency between two orbital states of the surface-binding electron is in the mm wave band. Therefore, with the above spin-orbit coupling,  the mm wave electric field can affect spin effectively, for example, results in its ac-Stark shift. During the driving for this effect, the orbital state of trapped electron is not excited by the mm wave signal.
Therefore, without the decoherence of orbital qubit, the mm wave electric field could be precisely measured by using the spin echo interferometry.

Practical experiments for implementing this interferometry would be complex, as the bias-current (for generating spin qubits) and the RF pulses (for manipulating qubits) are required to be precise enough. Given the fundamental limitations such as Johnson noise and the qubit decoherence are weaker than that within the usual systems, the present work showed still a promising platform for mm wave sensitive detection.
Hopefully, the presented circuit chip and ac-Stark effect as well as the spins detection may be also feasible for the system of electrons on solid neon~\cite{SchusterNature,Schuster2023,QCT}.

\section*{ACKNOWLEDGMENTS}
We thank Dr. Yiwen Wang for discussions. This work is partially supported by
the National Natural Science Foundation of China Grant No. 11974290, and the
National Key Research and Development Program of China under Grant No. 2021YFA0718803.

\setcounter{equation}{0}
\renewcommand\theequation{A\arabic{equation}}
\section*{APPENDIX A: THE ELECTRIC DIPOLE APPROXIMATION}
For a mm wave propagating along $x$-direction, its electromagnetic fields can be written as $\textbf{E}_{w}=\textbf{e}_zE_w\cos(\omega_T t-k_Tx)$ and $\textbf{B}_{w}=(\textbf{e}_x\times\textbf{E}_{w})/c$. As usual, $E_w$, $\omega_T$, $k_T$, and $c$ are the amplitude, frequency, wave number, and the propagating speed of the mm wave, respectively. The electromagnetic potentials of the mm wave are described by the fundamental equations  $\textbf{E}_w=-\nabla\phi_w-\partial_t\textbf{A}_w$ and $\textbf{B}_w=\nabla\times\textbf{A}_w$. Typically, we have $\phi_w=0$ and $\textbf{A}_w=-(E_w\textbf{e}_z/\omega_T)\sin(\omega_T t-k_Tx)$ under the Coulomb gauge $\nabla\cdot\textbf{A}_w=0$.

Using the presented solution of $\textbf{A}_w$, we perform a gauge transformation $\hat{U}_g=\exp[(ie/\hbar)\textbf{r}\cdot\textbf{A}_w]$ to Schr\"{o}dinger equation, then the Pauli Hamiltonian ({\color{blue}7}) becomes
\begin{equation}
\begin{aligned}
\hat{H}=&\hat{U}_g^\dagger\hat{H}_{\rm PL}\hat{U}_g
-i\hbar\hat{U}_g^\dagger \partial_t\hat{U}_g\\
=&\frac{(\hat{\textbf{p}}-e\textbf{A}')^2}{2m_e}+e\phi
+e\textbf{r}\cdot\partial_t\textbf{A}_w
+u_b\hat{\textbf{S}}\cdot\textbf{B}\,,
\end{aligned}
\end{equation}
with $\textbf{A}'=\textbf{A}_0-z(\textbf{e}_z\times \textbf{B}_w)$.
Note that, $\textbf{A}_0$ is the vector potential ({\color{blue}6}) of the applied static magnetic field, and $\phi=\Phi_0+\phi_w$ (with $\phi_w=0$) is the electrostatic potential for trapping electron on liquid helium.
Also, we note that the relation $\hat{U}_g^\dagger\hat{\textbf{p}}\hat{U}_g
=\hat{\textbf{p}}+e\textbf{A}_w+ez(\textbf{e}_z\times \textbf{B}_w)$ is used to exactly derive the second line in Eq.~({\color{blue} A1}).

Neglecting the $\textbf{A}'$-related terms can significantly simply the above Hamiltonian. This is a reasonable approximation, and called usually electric dipole approximation. Formally, Hamiltonian ({\color{blue}A1}) can be rewritten as
\begin{equation}
\begin{aligned}
\hat{H}&=\hat{H}_0-e\textbf{r}\cdot\textbf{E}_w
+u_b\hat{\textbf{S}}\cdot\textbf{B}+\hat{H}_{A'}\,,
\end{aligned}
\end{equation}
with $\hat{H}_0$ that has already been given by Eq.~({\color{blue}4}), and $\hat{H}_{A'}$ the Hamiltonian relating to $\textbf{A}'$. According Eq.~({\color{blue}6}) in the text and the magnetic field of mm wave introduced above, we can write $\hat{H}_{A'}$ approximately as
\begin{equation}
\begin{aligned}
\hat{H}_{A'}&\approx\frac{e B_0 z\hat{p}_y}{m_e}+\frac{e|\textbf{B}_w| z\hat{p}_x}{m_e}+\mathcal{O}(h_0^{-1})+\mathcal{O}(e^2)\,.
\end{aligned}
\end{equation}
Indeed, all the couplings presented in $\hat{H}_{A'}$ can be negligible, either because of the large detunings (between harmonic oscillators and the atomic qubit), or due to the small magnetic field $|\textbf{B}_w|=|\textbf{E}_w|/c$ of the very weak mm wave.
For the similar reason, we can write the term related to the electronic spin as
$u_b\hat{\textbf{S}}\cdot\textbf{B}=u_b\hat{\textbf{S}}\cdot(\textbf{B}_0+\textbf{B}_w)
\approx u_b\hat{\textbf{S}}\cdot\textbf{B}_0$ by neglecting the rapidly oscillating magnetic field $\textbf{B}_w$ of the weak mm wave. Therefore, we obtain Hamiltonian~({\color{blue}8}) presented in the text.

\setcounter{equation}{0}
\renewcommand\theequation{B\arabic{equation}}
\section*{APPENDIX B: THE SPIN-BOSON MODEL USED TO ESTIMATE COHERENCE TIMES}
In this proposal, a $y$-directional dc current $I_{\rm dc}$ beneath the liquid helium surface of depth $h_0$ is used to generate the desired spin-orbit coupling of a trapped electron.
Meanwhile, the current causes also decoherence into the spin superposition state due to the current fluctuation $\delta I$ and the height fluctuation $\delta h$ (capillary wave) of helium surface. With such fluctuations, the magnetic field $\textbf{B}_0$ in Eq.~({\color{blue}5}) is
corrected as
\begin{equation}
\begin{aligned}
\textbf{B}_{\text{c}}
\approx&\frac{\mu_0(I_{\rm dc}+\delta I)}{2\pi (h_0+ \delta h)}\left[\left(1-\frac{z}{h_0+\delta h}\right)\textbf{e}_x
-\frac{x}{h_0+\delta h}\textbf{e}_z\right]\\
\approx&\textbf{B}_{0}+B_0\left(\frac{\delta I}{I_{\rm dc}}-\frac{ \delta h}{h_0}\right)\textbf{e}_x\,,
\end{aligned}
\end{equation}
with $B_0=\mu_0I_{\rm dc}/(2\pi h_0)$, and where the high orders of small quantities, such as $z \delta h/h_0^2$ and $x \delta h/h_0^2$, are neglected.
Therefore, the noise reduces to the famous spin-boson model~\cite{Oxford1996,PRA1998,book2007,PRA2024}
\begin{equation}
\begin{aligned}
\hat{H}_{\rm sb}=\hbar\omega_0\left(\frac{\delta I}{I_{\rm dc}}
-\frac{\delta h}{h_0}\right)\hat{S}_z\,.
\end{aligned}
\end{equation}
Here, $\omega_0=\omega_s/2=\mu_bB_0/\hbar$ and $\omega_0/I_{\rm dc}=\mu_0\mu_b/(2\pi h_0\hbar )$, with $\omega_s$ being the transition frequency between two spin-states used in Eq.~({\color{blue}10}).

The spin-boson model explains decoherence by the following equation,
\begin{equation}
\begin{aligned}
\langle e^{i\phi_I(t)}\rangle\langle e^{i\phi_h(t)}\rangle
=e^{-\Gamma(t)}\,,
\end{aligned}
\end{equation}
with
\begin{equation}
\begin{aligned}
i\phi_I(t)=i\frac{\omega_0}{I_{\rm dc}}\left(\int_0^{T_1} \delta I {\rm d}t-
\int_{T_1}^{T_2} \delta I{\rm d}t\right)\,,
\end{aligned}
\end{equation}
and
\begin{equation}
\begin{aligned}
i\phi_h(t)=i\frac{\omega_0}{h_0}\left(\int_0^{T_1}\delta h {\rm d}t-
\int_{T_1}^{T_2}\delta h{\rm d}t\right)\,,
\end{aligned}
\end{equation}
due to the unitary evolution of Hamiltonian ({\color{blue}B2}) before any measurements performed to the system. Above, $T_1=t$ and $T_2=2t$ are the times of the two-stage ``free evolution" in the Hahn echo interferometry.

\subsection{The spin-boson model for Johnson-Nyquist noises}
Using the second quantization approach, the 1D current can be written as
\begin{equation}
\begin{aligned}
\delta I\approx i\sum_{k}
I_k[\hat{b}_k^\dagger e^{i(\omega_kt-ky)}-\hat{b}_ke^{-i(\omega_kt-ky)}]\,,
\end{aligned}
\end{equation}
with $I_k=\sqrt{\hbar\omega_k/(Ll_0)}$ and $\omega_k=|k|/\sqrt{l_0c_0}$ being the amplitude and the frequency of $k$-th boson mode, respectively. Note that, $k=\pm(1,2,3,\cdots)2\pi/L$, and $\hat{b}_k^\dagger$ and $\hat{b}_k$ are the corresponding creation and annihilation operators. $l_0$ is the inductance per unit length, $c_0$ is the capacitance per unit length, and $L$ a temporary length of the transmission line~\cite{Schoelkopf,RMP-circuit}.

As a consequence, the current-related Eq.~({\color{blue}B4}) is written as
\begin{equation}
\begin{aligned}
i\phi_I(t)=\sum_k\alpha_k\hat{b}_k^\dagger-\alpha_k^*\hat{b}_k\,,
\end{aligned}
\end{equation}
with
\begin{equation}
\begin{aligned}
\alpha_k=\frac{i \omega_0 I_k e^{-iky}}{I_{\rm dc}\omega_k}\left[\left(e^{i\omega_kt}-1\right)
-\left(e^{i2\omega_kt}-e^{i\omega_kt}\right)\right]\,.
\end{aligned}
\end{equation}
The spin-boson model~\cite{Oxford1996,PRA1998,book2007,PRA2024} says
\begin{equation}
\begin{aligned}
\langle e^{i\phi_I(t)}\rangle=\sum_k\langle e^{\alpha_k\hat{b}_k^\dagger-\alpha_k^*\hat{b}_k}\rangle=
e^{-\Gamma_I}\,,
\end{aligned}
\end{equation}
with
\begin{equation}
\begin{aligned}
\Gamma_I=\sum_k|\alpha_k|^2[\bar{n}(\omega_k)+\frac{1}{2}]\,,
\end{aligned}
\end{equation}
and $\bar{n}(\omega_k)=\langle\hat{b}_k^\dagger\hat{b}_k\rangle=[\exp(\hbar\omega_k/k_bT)-1]^{-1}$.

According to Eq.~({\color{blue}B8}), we have
\begin{equation}
\begin{aligned}
|\alpha_k|^2=\frac{4\omega_0^2 I_k^2}{I_{\rm dc}^2}
\left[\frac{\sin^2(\omega_kt/2)}{(\omega_k/2)^2}
-\frac{\sin^2(\omega_kt)}{\omega_k^2}\right]\,,
\end{aligned}
\end{equation}
and therefore,
\begin{equation}
\begin{aligned}
\Gamma_{I}=\frac{4\omega_0^2}{I_{\rm dc}^2}
\sum_kI_k^2[\bar{n}(\omega_k)+\frac{1}{2}]Y(\omega_kt)\,,
\end{aligned}
\end{equation}
with the time dependent function
\begin{equation}
\begin{aligned}
Y(\omega_kt)=\frac{\sin^2(\omega_kt/2)}{(\omega_k/2)^2}-
\frac{\sin^2(\omega_kt)}{\omega_k^2}\,.
\end{aligned}
\end{equation}

Now, we rewrite the quadratic amplitude as $I_k^2=\hbar\omega_k\omega_{\text {min}}/(2\pi R_0)$, with $\omega_{\text{min}}=2\pi/(L\sqrt{l_0c_0})$ and the impedance $R_0=\sqrt{l_0/c_0}$. As a consequence, for $L\rightarrow\infty$, Eq.~({\color{blue}B12}) can be expressed as an integral,
\begin{equation}
\begin{aligned}
\Gamma_{I}&=\frac{4\omega_0^2}{\pi I_{\rm dc}^2 R_0}\int_0^\infty\hbar\omega
[\bar{n}(\omega)+\frac{1}{2}]Y(\omega t)
{\text{d}}\omega\,.
\end{aligned}
\end{equation}
Note that, the following mathematical equation
\begin{equation}
\begin{aligned}
\int_{0}^{\infty}\frac{\sin^2(\tau\omega)}{(\pi \tau/2)\omega^2}\text{d}\omega=1
\end{aligned}
\end{equation}
holds for arbitrary $\tau$. Therefore, Eq.~({\color{blue}B14})
can be further written as
\begin{equation}
\begin{aligned}
\Gamma_{I}&=\frac{2\omega_0^2 t}{I_{\rm dc}^2 R_0}\int_0^\infty\hbar\omega
[\bar{n}(\omega)+\frac{1}{2}]\delta(\omega)
{\text{d}}\omega\,,
\end{aligned}
\end{equation}
with $t\rightarrow\infty$.

Furthermore, we note that, $\bar{n}(\omega)\approx0$ with $\hbar\omega\gg k_bT$, and $\bar{n}(\omega)\approx k_bT/(\hbar\omega)$ with $\hbar\omega\ll k_bT$. Thus, Eq.~({\color{blue}B16}) can be approximated solved as
\begin{equation}
\begin{aligned}
\Gamma_I
\approx t\frac{2k_bT}{ R_0 }\left(\frac{\omega_0^2}{I_{\rm dc}^2}\right)=
t\frac{2k_bT}{ R_0 }\left(\frac{\mu_0\mu_b}{2\pi h_0\hbar}\right)^2\,,
\end{aligned}
\end{equation}
with the aforementioned $\omega_0/I_{\rm dc}=\mu_0\mu_b/(2\pi h_0\hbar)$, and where $2k_bT/R_0$ refers to the Johnson-Nyquist current noise spectral~\cite{Schoelkopf}. The obtained result shows that a small impedance leads to the large decoherence.
Considering typically $R_0\approx\sqrt{\mu_0/\epsilon_0}\approx377\,\,\Omega$ of free space~\cite{Penning}, we have $\Gamma_I/t\approx9\times10^{-9}\,\,{\rm Hz}$ with $T=10\,\,{\rm mK}$ and $h_0=5\,\,\mu{\rm m}$. As it is expected~\cite{Lyon}, the spin decoherence due to the magnetic field of thermal current noise is negligible.

The voltage noise does not directly change the magnetic spin qubit, but can indirectly affect it. Note that, the detuning $\Delta_a$ between mm wave and the 1D hydrogen-like atom shall be perturbed by the electric fields noise due to the linear Stark effect of such a symmetry-breaking atom. Therefore, the inputting signal wave shall bring also decoherence into the spin interferometer. This decoherence can still be described by the spin-boson model, with the effective Hamiltonian $\hat{H}_{V}=\hbar\Omega_{V}\hat{S}_z$ of spin qubit. Here, $\Omega_{V}=4\Omega_{V}\Omega_{sz}\Delta_a/(\Delta_a^2-\Delta_s^2)$ describes the coupling strength, and $\Omega_{V}\approx e\kappa(z_{22}-z_{11})\delta V/(l\hbar)$ is the perturbing transition frequency of the atomic qubit within the Johnson-Nyquist voltage noise $\delta V$.

Corresponding to current boson model ({\color{blue}B6}), the voltage boson modes should be written as
\begin{equation}
\begin{aligned}
\delta V\approx \sum_{k}
V_k[\hat{b}_k^\dagger e^{i(\omega_kt-ky)}+\hat{b}_ke^{-i(\omega_kt-ky)}]\,,
\end{aligned}
\end{equation}
with $V_k=I_kR_0$ being the voltage amplitude of $k$-th boson mode. All other parameters  in the present equation are the same as that in Eq.~({\color{blue}B6}). Therefore, using the similar procedure, the spin decoherence due to the voltage noise is solved as
\begin{equation}
\begin{aligned}
\Gamma_V
= 2R_0k_bT\left[\frac{4\Omega_{sz}\Delta_a}{(\Delta_a^2-\Delta_s^2)}
\frac{e(z_{22}-z_{11})\kappa}{l\hbar}\right]^2 t\,,
\end{aligned}
\end{equation}
and here, $2R_0k_bT$ refers to the Johnson-Nyquist voltage noise spectral~\cite{Schoelkopf}.
Using the relation $(\Delta_a-\Delta_s)\approx\xi^2\Omega_{sz}$ in Sec.~IV-C, we have
\begin{equation}
\begin{aligned}
\Gamma_V
\approx2R_0k_bT\left[
\frac{2e(z_{22}-z_{11})\kappa}{\xi^2 l\hbar}\right]^2 t\,,
\end{aligned}
\end{equation}
In contrast to current noise ({\color{blue}B17}), the above voltage noise is proportional to impedance. Considering now a large impedance such as $R_0=10\,{\rm k\Omega}$, we have $\Gamma_V/t\approx0.73\,\,{\rm Hz}$ with the parameters $\xi=10$, $\kappa/l\approx0.22\times10^{5}\,{\rm m}^{-1}$, and $T=10\,\,{\rm mK}$ that are introduced in the text.

\subsection{The spin-boson model with superfluid capillary wave}
Similarly, the surface wave of liquid helium can be expressed as the 2D boson modes~\cite{Dykman-qubit}, i.e.,
\begin{equation}
\begin{aligned}
\delta h=\sum_{\textbf{k}}h_k[\hat{b}_\textbf{k}^\dagger
e^{i(\omega_kt-\textbf{k}\cdot\textbf{r})}
+\hat{b}_\textbf{k}
e^{-i(\omega_kt-\textbf{k}\cdot\textbf{r})}]\,,
\end{aligned}
\end{equation}
with $\textbf{k}=(k_x,k_y)$ being the 2D wave vector, and $k=\sqrt{\textbf{k}\cdot\textbf{k}}$ the corresponding wave number. $h_k$
and $\omega_{k}$ are respectively the amplitude and frequency of $\textbf{k}$-th wave. $\hat{b}_{\textbf{k}}^\dagger$ and $\hat{b}_{\textbf{k}}$ are the boson operators of this wave, obeying the commutation relation $[\hat{b}_{\textbf{k}'},\hat{b}_{\textbf{k}}^\dagger]
=\delta_{\textbf{k}'\textbf{k}}$.

Similar to ({\color{blue}B10}), the decoherence factor due to the capillary wave
is written as
\begin{equation}
\begin{aligned}
\Gamma_h=\sum_{\textbf{k}}|\beta_k|^2[\bar{n}(\omega_k)
+\frac{1}{2}]\,,
\end{aligned}
\end{equation}
with
\begin{equation}
\begin{aligned}
|\beta_k|^2=\frac{4\omega_0^2 h_k^2}{h_0^2}
\left[\frac{\sin^2(\omega_kt/2)}{(\omega_k/2)^2}
-\frac{\sin^2(\omega_kt)}{\omega_k^2}\right]\,.
\end{aligned}
\end{equation}

The ripple amplitude $h_k$ has already been given by many previous references, see, e.g.,~\cite{USSR}. It reads
\begin{equation}
\begin{aligned}
h^2_k=\frac{1}{2S_0}\frac{\hbar k\tanh(kh_0)}{\rho_0\omega_k}\,,
\end{aligned}
\end{equation}
with the nonlinear dispersion relation~\cite{USSR,PRB-continuum},
\begin{equation}
\begin{aligned}
\omega_k^2=(g_0+\frac{\sigma_0}{\rho_0}k^2)k \tanh(kh_0)\,.
\end{aligned}
\end{equation}
Above, $S_0$ is a temporary surface area of the liquid helium, $\sigma_0$ is the surface tension, $\rho_0$ is the helium density, and $g_0$ the effective gravitational acceleration.

Applying continuum approximation~\cite{PRB-continuum}, i.e.,
\begin{equation}
\begin{aligned}
\frac{1}{S_0}\sum_{\textbf{k}}
&\approx\frac{1}{(2\pi)^2}
\int_{-\infty}^{\infty}\int_{-\infty}^{\infty}\text{d}k_x\text{d}k_y\\
&
=\frac{1}{(2\pi)^2}\int_{0}^{\infty}\int_{0}^{2\pi}k\text{d}k\text{d}\theta\,,
\end{aligned}
\end{equation}
we have
\begin{equation}
\begin{aligned}
\Gamma_h=
\int_{0}^{\infty}f(\omega_k)\left[\frac{\sin^2(\omega_kt/2)}{(\omega_k/2)^2}
-\frac{\sin^2(\omega_kt)}{\omega_k^2}\right] {\rm d}k\,,
\end{aligned}
\end{equation}
with
\begin{equation}
\begin{aligned}
f(\omega_k)&=\left(\frac{\hbar\omega_0^2}{4\pi\rho_0 h_0^2}\right)\frac{k^2\tanh{(k h_0)}}
{\omega_k}[\bar{n}(\omega_k)
+\frac{1}{2}]\,.
\end{aligned}
\end{equation}
Unfortunately, due to the nonlinear dispersion relation ({\color{blue}B25}) of the capillary wave, we can not get a beautiful solution of $\Gamma_h$ liking ({\color{blue}B17}) derived by the transmission line model~\cite{Schoelkopf,RMP-circuit}.

\begin{figure}[tbp]
\includegraphics[width=7cm]{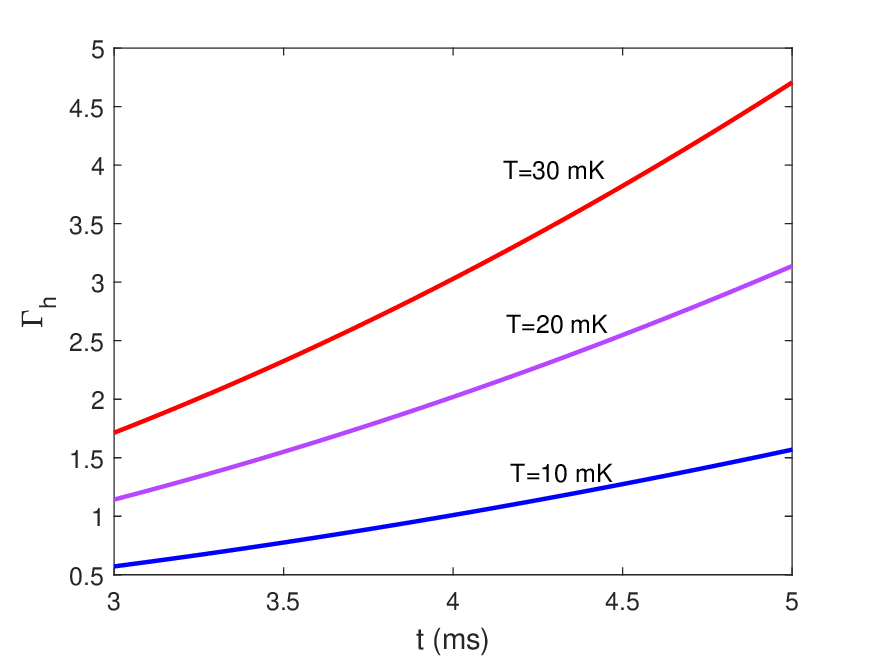}
\caption{The numerical solution of $\Gamma_h$. The results show that $\Gamma_h$ can be approximately regarded as a linear function of $t$, and the coherence time is about $4$~ms (with $T=10\,\,{\rm mK}$).}
\end{figure}

More directly, we use the numerical approach to compute the present decoherence factor $\Gamma_h$. For that, we rewrite ({\color{blue}B27}) as
\begin{equation}
\begin{aligned}
\Gamma_h=t\frac{\hbar\omega_0^2}{\sigma_0 h_0^2}\lim_{N \to \infty}\sum_{n=1}^{N}F(n)Y(n)\,,
\end{aligned}
\end{equation}
with
\begin{equation}
\begin{aligned}
F(n)=\frac{n^2\tanh(k_nh_0)}{(N\omega'_n)^3}[\bar{n}(\omega'_n)
+\frac{1}{2}]\,,
\end{aligned}
\end{equation}
and
\begin{equation}
\begin{aligned}
Y(n)=\frac{\sin^2(\omega'_nt'/2)}{\pi t'}
-\frac{\sin^2(\omega'_nt')}{4\pi t'}\,.
\end{aligned}
\end{equation}
Above, $\omega_n'=\omega_n\tau$ and $t'=t/\tau$ are respectively the dimensionless frequency and the dimensionless time, with $\tau=\hbar/(k_bT)\approx1\,\,{\rm ns}$ and $T=10\,\,{\rm mK}$.
The frequency is rewritten as $\omega_n=[(g_0+k_n^2\sigma_0/\rho_0)k_n \tanh(k_nh_0)]^{1/2}$, with the wave number $k_n=nk_{\rm max}/N$ and a cutoff of the maximal wave number $k_{\rm max}=(1/\tau)^2(\rho_0/\sigma_0)$. Such a cutoff of the wave number is defined by the maximal frequency $\omega_{\rm max}=1/\tau$ and the approximated dispersion relation $\omega_{\rm max}=\sqrt{\sigma_0k_{\rm max}^3/\rho_0}$ within the high-frequency regime~\cite{Dykman-qubit,Lamb-Shift}. It has been tested that the present high-frequency cutoff $\omega_{\rm max}=1/\tau$ is enough for the numerical computation of Eq.~({\color{blue}B29}).

Fig.~({\color{blue}6}) shows the results of $\Gamma_h$, which indicates that we can also express the damping factor as a linear function of time. The coherence time is estimated as $4\,\,{\rm ms}$ (with $T=10\,\,{\rm mK}$). The relevant parameters for the presented numerical computation are that, the surface tension of liquid helium is $\sigma_0\approx0.378\times10^{-3}\,\,{\rm N/m}$, the helium density is $\rho_0\approx0.154\times10^{3}\,\,{\rm kg}/{\rm m}^3$~\cite{Schuster2010}, the thickness of helium film is $h_0=5\,\,\mu{\rm m}$, and the gravitational acceleration is $g_0\approx9.8\,\,{\rm m}/{\rm s}^2$. Finally, we emphasize that the presented significantly larger decoherence than the above Johnson-Nyquist current noise is due to the large $\omega_0\approx0.2\,\,{\rm GHz}$ used to generate the strong spin-orbit coupling.

\setcounter{equation}{0}
\renewcommand\theequation{C\arabic{equation}}
\section*{APPENDIX C: TIME CORRELATION FUNCTIONS FOR THE SYMMETRY BREAKING TWO-LEVEL ATOM}
As mentioned in Sec.~III-A, the positional operator of a surface-state electron can be written as
\begin{equation}
\begin{aligned}
\hat{z}&=z_{11}\hat{\sigma}_{11}+z_{22}\hat{\sigma}_{22}+z_{12}\hat{\sigma}_{12}
+z_{21}\hat{\sigma}_{21}\\
&=\frac{z_{22}-z_{11}}{2}\hat{\sigma}_{z}+z_{12}\hat{\sigma}_{12}
+z_{21}\hat{\sigma}_{21}\,,
\end{aligned}
\end{equation}
within the two-level Hilbert space. In this space,
$\hat{\sigma}_{mn}=|m\rangle\langle n|$ is an operator (with $m=1,2$ and $n=1,2$) defined by the orbital eigenstates of electron, and $z_{mn}=\langle i|\hat{z}|j\rangle$ is the electric dipole matrix element. Specially, $\hat{\sigma}_{z}=|2\rangle\langle 2|-|1\rangle\langle 1|$ refers to the average-height difference between the two orbital states of electron floating on liquid helium.

In Heisenberg picture, the positional operator is formally written as
\begin{equation}
\begin{aligned}
\hat{z}(t)=&\frac{z_{22}-z_{11}}{2}\hat{\sigma}_{z}(t)+z_{12}\hat{\sigma}_{12}(t)
+z_{21}\hat{\sigma}_{21}(t)\,.
\end{aligned}
\end{equation}
The last two terms in this equation lead to the electric dipole radiation of mm waves. This is similar to the resonance fluorescence in the system of natural atoms. However, to detect such resonance fluorescence, one needs an additional mm wave sensor~\cite{Saturation}. Different from the natural atoms, the $\hat{\sigma}_{z}(t)$-term in Eq.~({\color{blue}C2}) is due to the symmetry breaking of electron floating on liquid helium, which can generate the observable low-frequency currents, as it is demonstrated by the experiment~\cite{DenisPRL2019}.

For an ac signal, Eq.~({\color{blue}24}) in Sec. IV-B can be rewritten as
\begin{equation}
\begin{aligned}
\hat{z}(\omega)
=\frac{z_{22}-z_{11}}{2}\hat{\sigma}_z(\omega)\,,
\end{aligned}
\end{equation}
with
\begin{equation}
\begin{aligned}
\hat{\sigma}_z(\omega)
=\frac{1}{T}\int_{0}^{T} \hat{\sigma}_{z}(t)\cos(\omega t){\rm d}t\,\,\,\,\,{\rm and}\,\,\,T\rightarrow\infty\,.
\end{aligned}
\end{equation}
This means that
\begin{equation}
\begin{aligned}
&\langle\hat{\sigma}_{z}(\omega)\hat{\sigma}_{z}(\omega)\rangle\\
&=\frac{1}{T^2}\int_{0}^{T}\int_{0}^{T}
\langle\hat{\sigma}_{z}(t_1) \hat{\sigma}_{z}(t_2)
\rangle\cos(\omega t_1)\cos(\omega t_2){\rm d}t_1{\rm d}t_2\,.
\end{aligned}
\end{equation}
Here, $\langle\hat{\sigma}_z(t_1)\hat{\sigma}_z(t_2)\rangle$ is the two-time correlation function of the signal, which can be evaluated by using the so-called quantum regression theorem~\cite{book2002}.

To use quantum regression theorem, we need to rewrite the above integral as
\begin{equation}
\begin{aligned}
&\langle\hat{\sigma}_{z}(\omega)\hat{\sigma}_{z}(\omega)\rangle\\
&=\frac{1}{T^2}\int_{0}^{T}\int_{0}^{T-t}\cos[\omega(t+\tau)]\cos(\omega t)
\\
&\,\,\,\,\,\,\,\,\,
\,\,\,\,\,\,\,\,\,\times[\langle\hat{\sigma}_{z}(t+\tau) \hat{\sigma}_{z}(t)\rangle+
\langle\hat{\sigma}_{z}(t) \hat{\sigma}_{z}(t+\tau)\rangle]{\rm d}\tau{\rm d}t\,.
\end{aligned}
\end{equation}
As it is expected, $\langle\hat{\sigma}_{z}(\omega)\hat{\sigma}_{z}(\omega)\rangle$ is real, because $\hat{\sigma}_{z}(t+\tau) \hat{\sigma}_{z}(t)+\hat{\sigma}_{z}(t) \hat{\sigma}_{z}(t+\tau)$ is a Hermitian operator.

In Sec.~IV-B, Eq.~({\color{blue}34}) is written in the form with atomic density matrix elements. Using the relations
$\rho_{21}(t)=\langle \hat{\sigma}_{12}(t)\rangle$, $\rho_{12}(t)=\langle \hat{\sigma}_{21}(t)\rangle$, and $\rho_{22}(t)=\langle \hat{\sigma}_{z}(t)\rangle/2+1/2$, we can reexpress Eq.~({\color{blue}34}) as
\begin{equation}
\begin{aligned}
&\frac{{\rm d}\langle\hat{\sigma}_z(t)\rangle}{{\rm d}t}
=2\Omega_{12}(t)\langle \hat{\sigma}_y(t)\rangle
-\Gamma_a\langle\hat{\sigma}_z(t)\rangle-\Gamma_a\,,\\
&\frac{{\rm d}\langle\hat{\sigma}_y(t)\rangle}{{\rm d}t}=-2\Omega_{12}(t)\langle\hat{\sigma}_z(t)\rangle
-\frac{\Gamma_a}{2}\langle\hat{\sigma}_y(t)\rangle\,,
\end{aligned}
\end{equation}
with $\hat{\sigma}_y(t)=i[\hat{\sigma}_{12}(t)-\hat{\sigma}_{21}(t)]$.
As mentioned in the text, Rabi frequency $\Omega_{12}(t)$ is time-dependent for the modulated strength of the driving field.

\begin{figure}
\subfigure[]{\includegraphics[width=6.5cm]{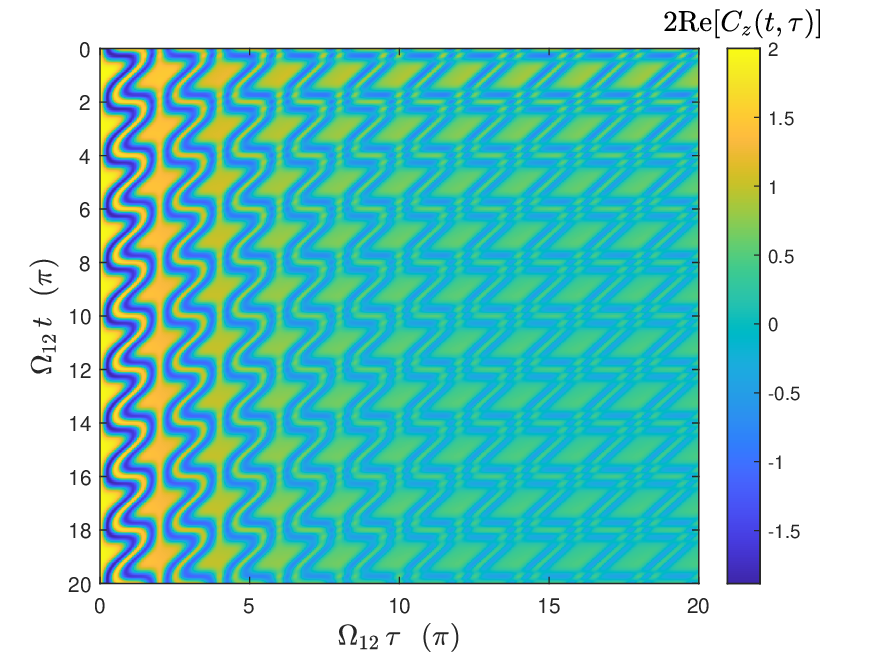}}
\subfigure[]{\includegraphics[width=6.5cm]{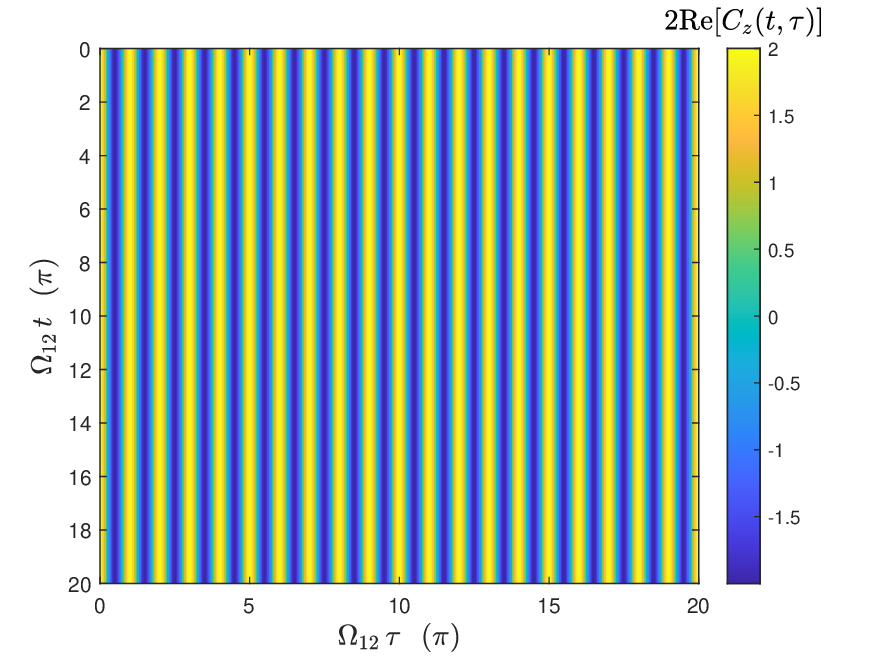}}
\caption{(a) The numerical solution $2{\rm Re}[C_z(t,\tau)]=\langle\hat{\sigma}_z(t)\hat{\sigma}_z(t+\tau)\rangle+{\rm c.c.}$ of Eqs.~({\color{blue}C8}) and ({\color{blue}C9}), with $\Omega_{12}(t)=[1+\cos(\omega_r t)]\Omega_{12}$ and $\omega_r=\Omega_{12}=10\,\Gamma_a$.
(b) To validate our computer program, we set $\Gamma_a=0$ and $\Omega_{12}(t)=\Omega_{12}$ to solve again Eqs.~({\color{blue}C8}) and ({\color{blue}C9}). With such hypothetical parameters, the analytical solution $\langle\hat{\sigma}_z(t)\hat{\sigma}_z(t+\tau)\rangle=\cos(2\tau\Omega_{12})$ is already known based on
Eq.~({\color{blue}C11}). As it is expected, Fig.~(b) agrees well with its analytical solution $2{\rm Re}[C_z(t,\tau)]=2\cos(2\tau\Omega_{12})$.  To plot Figs.~(a) and (b), we have assigned the initial values $\rho_{22}(t=0)=0$ and $\rho_{21}(t=0)=0$.}
\label{fig2}
\end{figure}

Supposing $\Gamma_a=0$ and removing out the measurement symbol
$\langle\cdots\rangle$, the above equations reduce to the standard Heisenberg equations. As a consequence~\cite{book2002}, we have
\begin{equation}
\begin{aligned}
\frac{\partial\langle\hat{\sigma}_z(t)\hat{\sigma}_z(t+\tau)\rangle}{\partial\tau}
=&2\Omega_{12}(t+\tau)\langle\hat{\sigma}_z(t)\hat{\sigma}_y(t+\tau)\rangle\\
&
-\Gamma_a\langle\hat{\sigma}_z(t)\hat{\sigma}_z(t+\tau)\rangle
-\Gamma_a\langle\hat{\sigma}_z(t)\rangle\,,\\
\end{aligned}
\end{equation}
and
\begin{equation}
\begin{aligned}
\frac{\partial\langle\hat{\sigma}_z(t)\hat{\sigma}_y(t+\tau)\rangle}{\partial\tau}
=&-2\Omega_{12}(t+\tau)\langle\hat{\sigma}_z(t)\hat{\sigma}_z(t+\tau)\rangle\\
&
-\frac{\Gamma_a}{2}\langle\hat{\sigma}_z(t)\hat{\sigma}_y(t+\tau)\rangle\,.
\end{aligned}
\end{equation}
Therefore, the desired function $\langle\hat{\sigma}_z(t)\hat{\sigma}_z(t+\tau)\rangle=C_z(t,\tau)$, as well as $\langle\hat{\sigma}_z(t)\hat{\sigma}_y(t+\tau)\rangle=C_y(t,\tau)$, is solvable with the ``initial" values:
\begin{equation}
\begin{aligned}
&C_{z}(t,0)=\langle\hat{\sigma}_z(t)\hat{\sigma}_z(t)\rangle=1\,,\\
&
C_{y}(t,0)=\langle\hat{\sigma}_z(t)\hat{\sigma}_y(t)\rangle=-i2{\rm Re}[\rho_{21}(t)]\,.
\end{aligned}
\end{equation}
Note that, Eqs.~({\color{blue}C8}) and ({\color{blue}C9})
do not violate the conjugate relations
$\partial_\tau\langle\hat{\sigma}_z(t)\hat{\sigma}_z(t+\tau)\rangle
=[\partial_\tau\langle\hat{\sigma}_z(t+\tau)\hat{\sigma}_z(t)\rangle]^*$ and $\partial_\tau\langle\hat{\sigma}_z(t)\hat{\sigma}_y(t+\tau)\rangle
=[\partial_\tau\langle\hat{\sigma}_z(t+\tau)\hat{\sigma}_y(t)\rangle]^*$.

Assuming again $\Gamma_a=0$ and setting $\Omega_{12}(t)=\Omega_{12}={\rm constant}$, Eqs.~({\color{blue}C8}) and ({\color{blue}C9}) reduce greatly to $\partial_\tau^2C_z(t,\tau)=-(2\Omega_{12})^2C_z(t,\tau)$.
This is nothing but the case of Rabi oscillation, and wherein the Heisenberg operators are well-known, for example,
\begin{equation}
\begin{aligned}
\hat{\sigma}_z(t)=&\cos(2\Omega_{12}t)\hat{\sigma}_z+\sin(2\Omega_{12} t)\hat{\sigma}_y\,.
\end{aligned}
\end{equation}
Inserting this solution into the correlation function $C_z(t,\tau)=\langle\hat{\sigma}_z(t)\hat{\sigma}_z(t+\tau)\rangle$, we can easily check that the mentioned equation $\partial_\tau^2C_z(t,\tau)=-(2\Omega_{12})^2C_z(t,\tau)$ holds really. This is nothing but a simple test for Eqs.~({\color{blue}C8}) and ({\color{blue}C9}).

\begin{figure}[tbp]
\includegraphics[width=7cm]{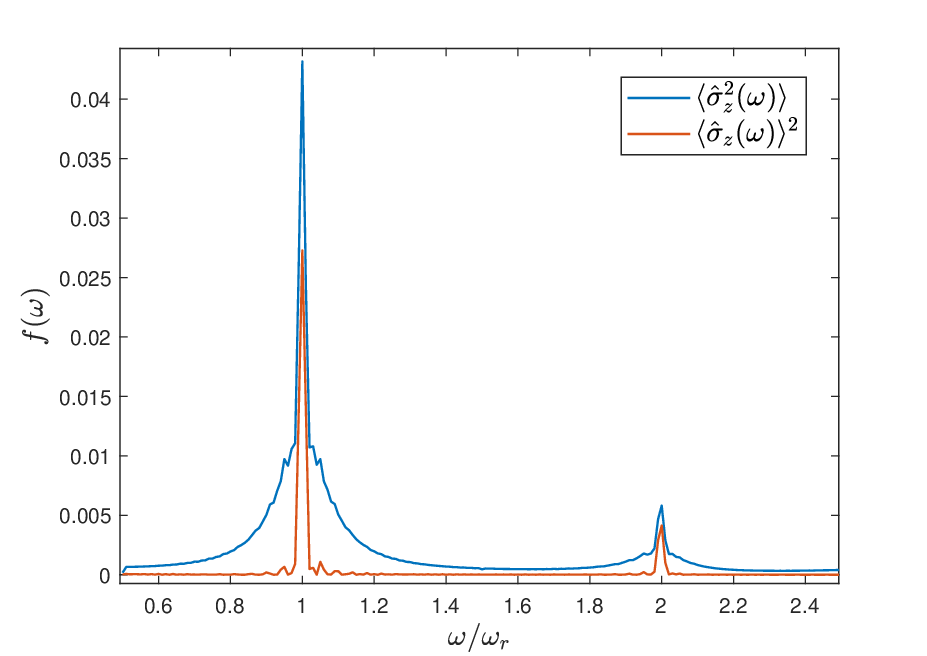}
\caption{(a) Numerical solutions of $f(\omega)=\langle\hat{\sigma}^2_z(\omega)\rangle$ (blue line) and $f(\omega)=\langle\hat{\sigma}_z(\omega)\rangle^2=[2\rho_{22}(\omega)]^2$ (red line), with $\Omega_{12}(t)=[1+\cos(\omega_r t)]\Omega_{12}$, $\omega_r=\Omega_{12}=10\,\Gamma_a$, and the initial values $\rho_{22}(t=0)=0$ and $\rho_{21}(t=0)=0$.}
\end{figure}

For our question, i.e., $\Gamma_a\neq0$ and $\Omega_{12}(t)\neq{\rm constant}$, analytically solving Eqs.~({\color{blue}C8}) and ({\color{blue}C9}) are almost impossible, but the numerical method works, as shown in Fig.~\textcolor{blue}{7}. To plot this figure, we need also the numerical solutions of the ``initial" values in Eq.~({\color{blue}C10}) as well as the term $\langle\hat{\sigma}_z(t)\rangle=2\rho_{22}(t)-1$ in Eq.~({\color{blue}C8}), wherein the density matrix elements $\rho_{21}(t)$ and $\rho_{22}(t)$ are described by Eq.~({\color{blue}34}) in the text.

With the numerical solution of Eqs.~({\color{blue}C8}) and ({\color{blue}C9}), integral ({\color{blue}C6}) can be calculated by using the following summation
\begin{equation}
\begin{aligned}
\langle\hat{\sigma}_{z}(\omega)\hat{\sigma}_{z}(\omega)\rangle
=&\frac{2}{N^2}\sum_{n=0}^{N}\sum_{m=0}^{N-n}{\rm Re}[C_z(n{\rm d}t,m{\rm d}\tau)]\\
&\times
\cos[\omega(n{\rm d}t+m{\rm d}\tau)]
\cos(\omega n{\rm d}t)\,,
\end{aligned}
\end{equation}
with $N\rightarrow\infty$ and ${\rm d}t={\rm d}\tau\rightarrow0$. The obtained results in Fig.~\textcolor{blue}{8} indicate that we can roughly treat $\langle\hat{\sigma}^2_z(\omega)\rangle\approx\langle\hat{\sigma}_z(\omega)\rangle^2$, and consequently use $\langle\hat{i}^2(\omega)\rangle=\langle\hat{i}(\omega)\rangle^2$ to evaluate the shot noise {(\color{blue}45}) in Sec.~IV-C.

\end{document}